\documentclass[11pt]{article}
\usepackage{amsmath,amssymb,bm,epsf,epsfig,graphicx}
\input epsf.sty
\usepackage[utf8]{inputenc}
\usepackage[english]{babel}
\usepackage{amsmath}
\usepackage{latexsym}
\usepackage{amsfonts}
\usepackage{mathtools}
\usepackage{amssymb}
\usepackage{scalerel}
\usepackage[left=3cm,right=3cm,top=3.1cm,bottom=3.1cm]{geometry}
\usepackage{cite}
\usepackage{tensor}
\usepackage{bbm}
\usepackage{xcolor}

\usepackage{hyperref}
\numberwithin{equation}{section}

    \def\ket#1{|#1 \rangle}
    \def\aver#1{\left\langle\, #1 \,\right\rangle}

    \def\p{\partial}

    \def \be {\begin{eqnarray}}
    \def \ee {\end{eqnarray}}
    \def \bal {\begin{align}}
    \def \eal {\end{align}}
    \def \bdm {\begin{displaymath}}
    \def \edm {\end{displaymath}}

    \def\del {\partial}
    \def\0{\nonumber}

    \def\w{\omega}

   \newcommand{\mycolor}{black}

\newenvironment{myblock}
  {\par\begingroup\color{\mycolor}\noindent}  
  {\par\endgroup}

\begin{document}
	\begingroup\allowdisplaybreaks

\vspace*{1.1cm}
\centerline{\Large \bf Gauge-invariant action for free string field theory with boundary}


\vspace{.3cm}

\begin{center}

{\large Carlo Maccaferri$^{(a)}$\footnote{Email: maccafer at gmail.com}, Alberto Ruffino$^{(a)}$\footnote{Email: ruffinoalb at gmail.com}  and  Jakub Vo\v{s}mera$^{(b)}$\footnote{Email: jakub.vosmera at ipht.fr} }
\vskip 1 cm
$^{(a)}${\it Dipartimento di Fisica, Universit\`a di Torino, \\INFN  Sezione di Torino \\
Via Pietro Giuria 1, I-10125 Torino, Italy}
\vskip .5 cm
\vskip .5 cm
$^{(b)}${\it Institut de Physique Théorique\\
	CNRS, CEA, Université Paris-Saclay\\
 Orme des Merisiers, Gif-sur-Yvette, 91191 CEDEX, France}

%
\end{center}

\vspace*{6.0ex}

\centerline{\bf Abstract}
\bigskip
We construct fully gauge-invariant kinetic terms for open and closed string field theories on a target space with boundary. This is realized by promoting the gauge parameters at the boundary to extra dynamical modes describing boundary degrees of freedom. Having a gauge-invariant classical action, we also construct the corresponding BV master action and show that the master equation is obeyed thanks to a peculiar nilpotent structure of the BV kinetic operator. From this general construction we explicitly derive free actions for massless and massive modes of open and closed strings propagating on a target with a boundary, including linear-dilaton backgrounds.

\baselineskip=16pt
\newpage
\setcounter{tocdepth}{2}
\tableofcontents

\section{Introduction and summary}\label{sec:intro}
There are several scenarios in string theory which feature non compact directions or boundaries, ranging from simple flat space to linear-dilaton,  Liouville, non-compact Calabi-Yau's, Anti-de-Sitter spaces and others.
Although these are well-known exact backgrounds  (when properly tensored with a complementary CFT so as to reach string criticality), the presence of non-compact directions can give rise to non-trivial infrared physics where the dynamical fields do not necessarily vanish at infinity or at the (conformal) boundary. In this case some of the basic assumptions which we generally take for granted (for example, the vanishing of BRST-exact world-sheet correlators) are no more true and we can get extra boundary contributions. It is fair to say that at the moment, we do not have a completely satisfactory worldsheet description of string theory in presence of these asymptotic effects. Part of the conceptual difficulty is that the non-local nature of the string seems to be at odds with the notion of having a boundary in spacetime. On a second thought however, this problem  deals with the long-distance infrared behaviour of string theory, which is expected to be qualitatively similar to that of a standard QFT containing massless excitations.  It seems therefore natural to approach this problem from the perspective of Sting Field Theory (SFT) (see \cite{Sen:2024nfd, Maccaferri:2023vns} for recent reviews and reference therein) which can be directly compared to other classical field theories like General Relativity (or Supergravity), where these `boundary' effects are taken into account by the Gibbons-Hawking-York (GHY) term \cite{York:1972sj, Gibbons:1976ue} which is also a key ingredient in the definition of the ADM observables \cite{Arnowitt:1962hi} of black holes.

In this paper we  provide a complete gauge-invariant construction of free open and closed bosonic SFT with target space $M\subset \mathbb{R}^D$,  having a  boundary $\del M$. As already anticipated in our previous paper \cite{paper1} (where we managed to write free gauge-invariant actions for the massless sector and the tensionless limit of string theory for a flat boundary), this is achieved by adding boundary terms containing new dynamical  boundary degrees of freedom. However the boundary fields we use in this paper are different from \cite{paper1} and they are essentially identified with the SFT gauge parameters, restricted to the boundary. As discussed in the introduction of \cite{paper1}, this is the very same mechanism that restores  gauge-invariance of Chern-Simons with boundary.\footnote{See  \cite{Balasubramanian:2018axm} for similar ideas on the entanglement entropy in string theory.}
Our action  thus contains the original bulk degrees of freedom in the string field $\Psi$ and new boundary degrees of freedom, assembled in a `boundary' gauge-string field $\chi$. The action is given by
\begin{align}
S_{\rm tot}(\Psi,\chi)=\frac12 \w_M\!\left(\Psi,Q\Psi\right)-\frac12 \w_{\del M}\!\left(\Psi,\Gamma^*\Psi\right)+\w_{\del M}\!\left(\Psi,\Gamma^*Q\chi\right)-\frac12 \w_{\del M}\!\left(Q\chi,\Gamma^*Q\chi\right)\,,\label{Stot-gen}
\end{align}
where the grassmann-odd operator $\Gamma^*$ is linear in the normal derivative to the boundary $\del_\perp$ and is defined in (\ref{B-Gamma},\ref{eq:GammaDecomp}).   Here $\w_M$ is the symplectic form (BPZ inner product) restricted to $M\subset\mathbb{R}^D$ \eqref{w-M} and $\w_{\del M}$ is its further restriction to the boundary $\del M$ \eqref{w-del}. In particular the last three terms are pure boundary terms which couple the bulk string field $\Psi$ with the boundary degrees of freedom contained in $\chi$. This action has the following consistency properties
\begin{enumerate}
\item It has a well-defined variational principle in the sense discussed in \cite{paper1}. 
This is achieved thanks to the first $\Psi$-$\Psi$ boundary term which renders the $\Psi$-action  first order in derivatives, thus giving a well-defined variational principle allowing for Dirichlet or Neumann-like boundary conditions. Note that an equivalent $\Psi$-$\Psi$ term, which makes the variational principle well-defined, was first constructed by \cite{georg, atakan}.
\item It is invariant under the following gauge transformations
\begin{subequations}
\begin{align}
\delta_g\Psi&=Q\Lambda\,,\\
\delta_g\chi&=\Lambda+Q\Upsilon\,.
\end{align}
\end{subequations}
Notice that the gauge transformation of $\Psi$ is the same as in the boundary-less case, which means  that   the boundary is now compatible with the full gauge-symmetry of the boundary-less theory, without any restriction on the bulk field. As anticipated, and as explicitly visible from these gauge transformations, the boundary field $\chi$ lives in the same space as the gauge parameter $\Lambda$ and  it inherits from it the reducible structure given by the extra `gauge for gauge' parameter $\Upsilon$. In particular, in both the open and closed string cases, $\Psi$ and $\chi$ have opposite grassmannality. This gauge-invariance is valid for the full tensile string theory and thus provides an important upgrade wrt to \cite{paper1}.
\item It reduces to the standard $\frac12\w(\Psi,Q\Psi)$ action when the boundary is sent to $\infty$ and $\Psi$ has sufficient fall-off conditions.
\end{enumerate}
This is the main result of the paper, which is organized as follows. In section \ref{sec:2} we build the gauge-invariant action and consequently the BV master action. Interestingly the master action is not simply given by un-restricting the ghost number in  \eqref{Stot-gen} but it displays a richer structure. Nonetheless we show that  the classical master equation follows from a nilpotency condition of the full BV kinetic operator, which generalizes the standard $Q^2=0$ of the boundary-less case. In section \ref{sec:examples} we give explicit realizations of \eqref{Stot-gen}  on the half-space $M=\mathbb{R}_+\times \mathbb{R}^{D-1}$ for the massless and the first massive level of the open string, as well as for the massless level of the closed string. In the latter case, after integrating out some auxiliary fields, we find perfect agreement with the standard Kalb-Ramond-dilaton gravity supplemented with GHY term \eqref{eq:GRaction}, expanded around flat space with a flat boundary. In particular the dynamical boundary modes in this case correspond to the transverse fluctuations of the boundary, analogously to what has been found in \cite{paper1}. In subsection \eqref{subsec:lin_dilaton} we generalize our construction to the linear-dilaton background on the half-
space. This is preceded by the crucial realization (which, to the best of our knowledge, has not been discussed elsewhere) that dilaton-gravity with boundary needs a boundary cosmological constant (in addiction to the well-known bulk one) in order for the linear-dilaton background to be a true stationary point without boundary tadpoles. After this important clarification, we find perfect agreement between the expanded gravity action and the effective action obtained from our SFT construction and we observe how the dilaton gradient  $\beta^\mu$  modifies the structure of the boundary action as compared to the $\beta^\mu=0$ case.  In  section  \ref{sec:disc} we conclude with a couple of open questions and an outlook for the future.
In appendix \ref{app:equiv} we give two notable alternative forms of the action and show how they are related to our main proposal \eqref{Stot}.

\section{Gauge-invariant action}\label{sec:2}
In this section we give the detailed derivation of our gauge-invariant action, together with the construction of the corresponding BV master action.

\subsection{Preliminaries and recap from \cite{paper1}}

We start with a bosonic closed string background containing $D$ non-compact free bosons describing $\mathbb{R}^D$. For definiteness we work in Euclidean space but our main results trivially extend to Minkowski \footnote {See subsection \ref{subsec:lin_dilaton} for the slight changes introduced by a linear dilaton background.}. The remaining part of the matter sector can be a generic  CFT with $c=26-D$ . The resulting string effective action on $\mathbb{R}^D$ will be schematically given by
\begin{align}
S(\phi_i)=\int_{\mathbb{R}^D}\,d^D x \,L(\phi_i, \del\phi_i, \del^2\phi_i,\cdots)\,,
\end{align}
where the $\phi^i(x)$ are the generic $\mathbb{R}^D$ space-time fields created by the string degrees of freedom with $x$ identified with the string center of mass.
To be precise we will find it useful to express the dynamical (open or closed) string field in position space as
\begin{align}
\ket\Psi=\Psi(x)\ket0\,,
\end{align}
where $x^\mu$ is the center of mass position operator, $p_\mu$ is its conjugate momentum
\begin{align}
[x^\mu,p_\nu]&=i\delta^\mu_\nu
\end{align}
and $\ket0$ is the zero-momentum $SL_2$-invariant vacuum. 
We can then define the free (open or closed) SFT as
\begin{align}
S[\Psi]=\frac12\,\w(\Psi, Q\Psi)=\frac12\int_{\mathbb{R}^D}\,d^D x \,\w'(\Psi(x), Q\Psi(x))\,.\label{actionRD}
\end{align}
The symplectic form $\w(\cdot,\cdot)$ is related to the BPZ inner product $\langle\cdot,\cdot\rangle$ as
\begin{align}
\w(\psi_1,\psi_2)=(-1)^{d_{\psi_1}}\aver{\psi_1,\#\,\psi_2}\,,
\end{align}
where $d_\psi$ is the degree, which is ghost number minus two for closed strings and ghost number minus one for open strings. Moreover we have $\#=c_0^-=\frac12(c_0-\bar c_0)$ for closed strings and $\#=1$ for open strings.  The reduced symplectic form $\omega'$ performs the inner product without integrating over the position zero modes.
The  (open or closed) BRST charge $Q$  can be written as
\begin{align}
Q=-c\del^\mu\del_\mu-\Omega^\mu\del_\mu+ Q'\label{Qder}\,,
\end{align}
where we have defined
\begin{subequations}
\begin{align}
c&=\alpha' c_0\,,\\
\Omega^\mu&=\,i\sqrt{2\alpha'}\sum_{n\neq 0}c_{-n} \alpha^\mu_n\,,
\end{align}
\end{subequations}
for the open string and
\begin{subequations}
\begin{align}
c&=\frac{\alpha'}{4}  (c_0+\bar c_0)\equiv \frac{\alpha'}{2}c_0^+\,, \\
\Omega^\mu&=\,i\sqrt{\frac{\alpha'}{2}}\sum_{n\neq 0}\left(c_{-n} \alpha^\mu_n+\bar c_{-n} \bar\alpha^\mu_n\right)\,,
\end{align}
\end{subequations}
for the closed string. Finally, $Q'$ is free of $\mathbb{R}^D$ derivatives and it is the needed completion to the full BRST charge in the open or closed Hilbert space (the closed Hilbert space is, as usual, annihilated by $b_0^-$ and $L_0^-$).

Assuming sufficient fall-off conditions for $\Psi(x)$ towards infinity, we can verify that $Q$ is a BPZ-odd operator $(Q^*=-Q)$  in the sense that \footnote{The notion of bpz conjugation is always understood wrt the $\mathbb{R}^D$ symplectic form $\w(\cdot,\cdot)$ with assumed fall-off conditions. }
\begin{align}
\w(\psi_1,Q \psi_2)=(-1)^{d_{\psi_1}+1}\w(Q\psi_1, \psi_2)\,.
\end{align}
From this it follows that the free SFT action on $\mathbb{R}^D$ (with sufficient fall-off conditions)  has the expected variation
\begin{align}
\delta S=\w(\delta\Psi,Q\Psi)
\end{align}
and the expected gauge invariance
\begin{subequations}
\begin{align}
\delta_g \Psi&=Q\Lambda\,,\\
\delta_g S&=\w(Q\Lambda,Q\Psi)=\w(\Lambda,Q^2\Psi)=0\,.
\end{align}
\end{subequations}
Now we want to define the analog of \eqref{actionRD} on a subset $M\subset \mathbb{R}^D$ with a smooth boundary $\partial M$, without constraining in any way the  behaviour of the string field $\Psi(x)$ at the boundary. The obvious starting point is to  substitute 
$$
\int_{\mathbb{R}^D}\,d^D x\longrightarrow \int_{M}\,d^D x\,,$$
which, at the level of the $\mathbb{R}^D$-symplectic form $\w$ means \cite{Witten:1986qs,Cho:2023khj}
\begin{align}
\w(\cdot,\cdot)\longrightarrow\w_M(\cdot,\cdot)\equiv  \w(\cdot, \Theta\,\cdot)\,,\label{w-M}
\end{align}
where $\Theta=\Theta(x)$ is the characteristic function of $M$
\begin{align}
\Theta(x) =
\begin{cases}
1\,, & x \in M \,,\\
0\,, &  \text{otherwise} \,.
\end{cases}
\end{align}
With this modification  the action becomes
\begin{align}
\tilde S_0[\Psi]=\frac12\,\w(\Psi, \Theta Q\Psi)=\frac12\int_{M}\,d^D x \,\w'(\Psi(x), Q\Psi(x))\,.\label{actionM}
\end{align}
However since $Q$ is a second-order differential operator and the field $\Psi(x)$ is not restricted at the boundary $\del M$, the variational principle is not well-defined  as described in detail in \cite{paper1, georg, atakan}. Moreover the action \eqref{actionM} is also not gauge-invariant due to non-vanishing boundary terms in the variation.

\subsection{Construction of a consistent gauge-invariant quadratic action}
To construct a gauge-invariant action for a free string which constitutes a well-posed variational principle, we proceed in three steps.

\subsubsection{ Bulk action}
To start with, following  \cite{paper1},  we write an action where only first-order $\del_\mu$-derivatives are present, so that the variational principle is well-defined. This action is given by
\begin{align}
S_0(\Psi)=\frac12 \w(\Psi, (\Theta Q-\delta \Gamma^*)\Psi)\,.
\end{align}
Here the $\delta=\delta(x)$ distribution-like operator is the Dirac delta localizing at the boundary $\del M$
\begin{align}
\w_{\del M}(\psi_1,\psi_2)=\w(\psi_1,\delta\,\psi_2)\equiv\int_{\del M}d^{D-1}y\,\sqrt{h}\,\w'(\psi_1(x(y)),\psi_2(x(y)))\,,\label{w-del}
\end{align}
where we have introduced the induced metric on the boundary  
\begin{subequations}
\begin{align}
h_{ab}&\equiv \partial_a x^\mu(y)\,\partial_b x^\nu(y) \,\eta_{\mu\nu}\,,\\
h&\equiv\det(h_{ab})\,,
\end{align}
\end{subequations}
and $x^\mu(y)$ are embedding maps from general local boundary coordinates $y^a$ to the bulk flat coordinates $x^\mu$. 

The operator $\Gamma^*$ is defined as follows. Consider first the distribution-like operator
\begin{align}
B\equiv [\Theta, Q]\,.
\end{align}
This operator quantifies the failure of integrating by part $Q$ because of the presence of the boundary $\del M$ (which was noticed already in \cite{Kraus})
\begin{align}
\omega_\Theta(\psi_1, Q \psi_2)&=\omega(\psi_1,\Theta Q \psi_2)=(-1)^{d_1+1}\omega(Q\psi_1,\Theta \psi_2)+\omega(\psi_1,B \psi_2)\0\\
&=(-1)^{d_1+1}\omega_\Theta(Q\psi_1, \psi_2)+\omega(\psi_1,B \psi_2)\,.
\end{align}
Evaluating the commutator using \eqref{Qder} we find that indeed the distribution-like operator $B$ has the effect of localizing at the boundary 
\begin{align}
    B\,\psi = c\big[\delta_\mu(x)\p^\mu\!\psi +\p^\mu(\delta_\mu(x)\,\psi)\big] + \Omega^\mu \delta_\mu \,\psi \,,
\end{align}
where we have defined
\begin{align}
    \delta_\mu(x) \equiv \p_\mu \Theta(x) = -n_\mu(y)\,\delta(x)\,,
\end{align}
with $n_\mu(y)$ being the globally defined normal unit vector of the boundary $\del M$, pointing outside $M$.
The first-order differential operator $\Gamma^*$ is defined through the identification
\begin{align}
B\equiv \delta\Gamma^*+\Gamma\delta\,,\label{B-Gamma}
\end{align}
which gives
\begin{align}
    \Gamma^\ast = -c\, n^\mu(y)\,\p_\mu -\gamma^\ast=c\del_\perp-\gamma^\ast,\label{eq:GammaDecomp}
\end{align}
where we have defined the normal derivative $ \del_\perp\equiv -n^\mu(y)\,\p_\mu$, which points to the interior of $M$, and where $\gamma^*$ is any operator which is free of normal derivatives (so that it commutes with $\delta$) and  satisfies
\begin{align}
    \gamma^* +  \gamma= \Omega^\mu n_\mu(y)\,.\label{eq:gammaDecomp}
\end{align}
Making a special choice of $\gamma$ such that $\gamma=\gamma^*=\frac{1}{2}\Omega^\mu n_\mu(y)$, one would recover the boundary term which was first discussed by \cite{georg,atakan}. 
Notice that the full kinetic operator is now cyclic and BPZ odd
\begin{align}
(\Theta Q-\delta\Gamma^*)^*=Q^*\Theta-\Gamma\delta=-Q\Theta-\Gamma\delta=-\Theta Q +B-\Gamma\delta=-(\Theta Q- \delta \Gamma^*)\,.
\end{align}
Explicitly we find
\begin{align}
    S_0(\Psi)&=\frac{1}{2}\int_{M}d^d x\, \omega'\big(\p_\mu\Psi(x), c\p^\mu\Psi(x)\big)+\frac{1}{2} \int_{M}d^d x\, \omega'\big(\Psi(x),(Q'-\Omega^\mu\p_\mu)\Psi(x)\big)+\nonumber\\
    &\hspace{6.6cm}+\frac{1}{2}\int_{\p M}d^{d-1} y\,  \sqrt{h}\,\omega'\big(\Psi(y),\gamma^\ast \Psi(y)\big)\,.\label{eq:explicitS}
\end{align}
This action has now only first derivatives and so the variational principle is well-defined. 

The action $S_0$ gives a well-defined variational principle for any choice of $\Gamma^*$. However, as discussed in \cite{paper1}, its gauge variation is not zero and it is localized at the boundary. Namely, under the variation $\delta_g\Psi=\,Q\Lambda$, we obtain
\begin{align}
\delta_g S_0=-\w(\Psi,\delta\Gamma^* Q\Lambda)=&-\int_{\del M} d^{d-1} y\, \sqrt{h}\,\omega'\big(\Psi(x(y)),\Gamma^*Q\Lambda(x(y)))\,.\label{gauge-fail}
\end{align}
Notice that this total gauge variation is sensible to the gauge parameter $\Lambda$ at the boundary up to its second normal derivative $\del_\perp^2$ . In particular the third derivative contribution potentially contained in $\Gamma^*Q$ is accompanied by  $c^2=0$ and is therefore not present.

\subsubsection{Bulk-boundary action}
As already discussed in our previous paper \cite{paper1} it is in general not possible to choose  $\gamma$ in \eqref{eq:GammaDecomp} such that $\delta\Gamma^*Q\Lambda=0$. 

To rescue the part of gauge invariance which has been broken by the boundary, we need to add boundary degrees of freedom whose gauge variation can compensate for \eqref{gauge-fail}. In our previous paper \cite{paper1} we partially realized this idea by adding extra boundary string fields $\Sigma_i(y)$ sharing the same ghost-number/degree of the bulk field $\Psi$ and further constraining the $\Sigma$'s with a linear constraint. In this way we managed to define gauge invariant actions for the massless sector and  the tensionless limit of the full string theory.  Here, to make the full tensile string theory gauge invariant, we follow a different strategy that is directly inspired by the analogous problem  of abelian Chern-Simons with a boundary (see the introduction of \cite{paper1}).
 Looking at \eqref{gauge-fail} we see that this can be canceled by promoting the gauge parameter $\Lambda(x)$ approaching the boundary to a `boundary field' $\chi(x)$  naturally subject to the gauge transformation
\begin{align}
\delta_g\,\chi=\Lambda\,.
\end{align} 
With this preparation it is immediate to realize that \eqref{gauge-fail} can be canceled by a bulk-boundary action
\begin{align}
S_1(\Psi,\chi)=\w(\Psi,\delta\Gamma^* Q\chi)=\int_{\del M} d^{d-1} y\, \sqrt{h}\,\omega'\big(\Psi(x(y)),\Gamma^*Q\chi(x(y)))
\end{align}
when we vary with respect to $\chi$.  Notice that just as \eqref{gauge-fail} only depends on the values of $\Lambda(x)$, $\del_\perp\Lambda(x)$, $\del_\perp^2\Lambda(x)$ at the boundary $x\in\del M$, the bulk-boundary action $S_1(\Psi, \chi)$ only depends on 
\begin{subequations}
\label{eq:chis}
\begin{align}
\chi_{-1}(y)&\equiv \chi(x){\Big|}_{x\in\del M}\,,\\
\chi_{0}(y)&\equiv \del_\perp\chi(x){\Big|}_{x\in\del M}\,, \\
\chi_{1}(y)&\equiv \del_\perp^2\chi(x){\Big|}_{x\in\del M}\, .
\end{align}
\end{subequations}
The indices $(-1,0,1)$ on $\chi$ are (mass) space-time dimensions which are defined so that the bulk field $\Psi$ has conventionally dimension 1 and $\del_\mu$ has also dimension 1. This means that the BRST charge $Q$ has total mass-dimension two\footnote{Here we are ignoring the $(-2)$ mass dimensions of $\alpha'$} and therefore the gauge parameter $\Lambda$ has dimension $-1$, which is also inherited by $\chi$. With these conventions the action then needs total mass dimension 4. When we introduce the boundary,  the distribution $\Theta$ is  dimensionless and $\delta\sim\del\Theta$ has dimension 1.
Although the fields $\chi_{-1,0,1}(y)$ are the only boundary degrees of freedom which have been truly added,  we will still find it convenient to `package' them in  $\chi(x)$, having in mind that the action is only sensitive to the behaviour of $\chi(x)$ infinitesimally close to the boundary.

Notice importantly  that the  string field $\chi$ has degree $-1$ which means it is ghost-number 1 for closed strings and ghost-number 0 for open strings, just like the gauge parameter $\Lambda$. In particular its grassmannality is the opposite of the bulk field $\Psi$.

\subsubsection{Boundary action}
By construction the $\chi$ variation of $S_1$ cancels \eqref{gauge-fail}, but this is not sufficient  because $S_1$ also depends on $\Psi$, whose gauge variation is still unbalanced. Explicitly we find
\begin{align}
\delta_g(S_0+S_1)=\w(Q\Lambda,\delta\Gamma^* Q\chi)\,.
\end{align}
This variation can be easily canceled by the pure boundary action
\begin{align}
S_2(\chi)=-\frac12\w\left(Q\chi,\delta\Gamma^*\,Q\chi\right)=-\frac12\w\left(\chi,Q\delta\Gamma^*\,Q\chi\right)\,.
\end{align}
Notice that the ghost-number three `kinetic operator' $Q\delta\Gamma^*\,Q$ is BPZ even
\begin{align}
(Q\delta\Gamma^*\,Q)^*=-Q^*\Gamma\delta Q^*=-Q\Gamma\delta Q=Q(-B+\delta\Gamma^*)Q=Q\delta\Gamma^*\,Q\,,
\end{align}
thanks to $Q^2=0$ and 
\begin{align}
[Q,B]=[Q,[\Theta,Q]]=\frac12[\Theta,[Q,Q]]=0\,.
\end{align}
Computing the total gauge variation and using the even BPZ parity of $Q\delta\Gamma^*\,Q$ we then immediately find
\begin{align}
\delta_g(S_0+S_1+S_2)=\w(Q\Lambda,\delta\Gamma^* Q\chi)-\w\left(Q\Lambda,\delta\Gamma^*\,Q\chi\right)=0\,.
\end{align}
To summarize: we have found the following bulk-boundary gauge-invariant action
\begin{align}
S_{\rm tot}(\Psi,\chi)=\frac12 \w(\Psi, (\Theta Q-\delta \Gamma^*)\Psi)+\w(\Psi,\delta\Gamma^* Q\chi)-\frac12\w\left(Q\chi,\delta\Gamma^*\,Q\chi\right)\,.\label{Stot}
\end{align}
Notice that $\chi$ only appears in the combination $Q\chi$ which means that it is defined up to $Q$-exact terms. Therefore the action is invariant under the following gauge transformation
\begin{subequations}
\label{eq:gauge_both}
\begin{align}
\delta_g\Psi&=Q\Lambda\,,\label{gauge1}\\
\delta_g\chi&=\Lambda+Q\Upsilon\,,\label{gauge2}
\end{align} 
\end{subequations}
where $\Upsilon$ is a degree $-2$ (effectively grassmann even) extra gauge parameter, which is simply inherited from the reducibility of the $\Lambda$-gauge transformation and the fact the $\chi$ is morally the same as $\Lambda$ at the boundary.

Finally, let us notice that the action \eqref{Stot} can be equivalently expressed as
\begin{align}
    S_{\rm tot}(\Psi,\chi)=\frac12 \w\big((\Psi-Q\chi), (\Theta Q-\delta \Gamma^*)(\Psi-Q\chi)\big)\,\label{eq:StotGI},
\end{align}
where gauge-invariance trivially follows from the fact that $(\Psi-Q\chi)$ is itself gauge-invariant.
Here it is no longer manifest that $\chi$ only contributes through boundary terms: in order to write \eqref{eq:StotGI} down, one formally needs to extend $\chi$ into the bulk of $M$ and notice that the possible bulk terms where $\chi$ would appear are vanishing because of $Q^2=0$ and $\Theta Q-\delta\Gamma^*=Q\Theta + \Gamma\delta$. This identity can be also used to show that the boundary part of the action is free from normal derivatives, which is no more apparent at the level of \eqref{eq:StotGI}. 


\subsubsection{Super-matrix structure}
It is convenient to collect the dynamical classical fields $\Psi$ and $\chi$ in the ``super vector''
\begin{align}
\boldsymbol{\Psi}= \left(\begin{matrix}\Psi\\\chi \end{matrix}\right)=\left(\begin{matrix}\textrm{bulk mode}\\\textrm{boundary mode} \end{matrix}\right)\,,
\end{align}
where the lower entry has one degree less than the upper one and thus opposite grassmannality.  We define the (super) degree of $\boldsymbol{\Psi}$ to be the degree of its top component $\Psi$. In this notation we can simply write
\begin{align}
S_{\rm tot}=\frac12 \w\left(\boldsymbol{\Psi}^T, {\bf Q}\boldsymbol{\Psi}\right)\,,\label{Stotbf}
\end{align}
where we have introduced the  super matrix
\begin{align}
{\bf Q}=\left(\begin{matrix}\Theta Q-\delta\Gamma^*&\delta\Gamma^*Q \\ -Q\Gamma\delta&-Q\delta\Gamma^*Q \end{matrix}\right)\,.
\end{align}
To discuss the BPZ property of this distribution-like operator,  we have first of all to specify what we mean by (super)-transposition.  In our conventions the super-transpose  $(\cdot)^T$   is standard transposition for  super-vectors 
\begin{align}
 \left(\begin{matrix}\Psi\\\chi \end{matrix}\right)^T\equiv(\Psi,\chi)\,,
\end{align}
while on a $(2\times 2)$ supermatrix  is defined as 
\begin{align}
\left(\begin{matrix}a&b\\c&d\end{matrix}\right)^T\equiv\left(\begin{matrix}a&(-1)^{d_c }c\\b&(-1)^{d_d}d\end{matrix}\right)\,, 
\end{align}
where $d_{(\cdot)}$ is  the degree (which is  effectively the grassmanality). This definition ensures that given two even supervectors $(v,w)$ and a  supermatrix $A$ we have
\begin{align}
v^T A w=(A^T v)^T w\,.
\end{align}
 Given this definition, we can easily check that ${\bf Q}$ is super-BPZ odd in the sense that
\begin{align}
{\bf Q}^{\star}\equiv{\bf Q}^{*T}=\left(\begin{matrix}-(\Theta Q-\delta\Gamma^*)&-\delta\Gamma^*Q \\ Q\Gamma\delta&Q\delta\Gamma^*Q \end{matrix}\right)=-{\bf Q}\,.
\end{align}
In this notation the gauge invariance (\ref{gauge1},\ref{gauge2}) is expressed as
\begin{align}
\delta_g \boldsymbol{\Psi}={\bf G}\boldsymbol{\Lambda}\,,\label{boldgauge}
\end{align}
where
\begin{align}
\boldsymbol{\Lambda}=&\left(\begin{matrix}\Lambda \\ \Upsilon \end{matrix}\right),\\
{\bf G}=&\left(\begin{matrix} Q&0 \\ 1&Q \end{matrix}\right)\,.
\end{align}
According to the convention described above, the super-BPZ conjugate of ${\bf G}$ is
\begin{align}
{\bf G}^\star={\bf G}^{*T}=\left(\begin{matrix} -Q&1 \\ 0&Q \end{matrix}\right)\,.
\end{align}
With this preparation, the gauge variation of the action is given by
\begin{align}
\delta_g S_{\rm tot}=\w\left(\boldsymbol{\Psi}^T, {\bf Q} {\bf G}\boldsymbol{\Lambda}\right)=0\,,
\end{align}
because, as it easy to check
\begin{align}
 {\bf Q} {\bf G}=0\,.
\end{align}
Notice that ${\bf Q}$ is not nilpotent and, being a distribution-like operator containing $\delta$, ${\bf Q}^2$ is actually ill-defined. However, as we have just seen this is not a problem  since this product does not appear in the proof of gauge invariance.
On the other hand the non-distribution-like operator ${\bf G}$ {\it is} nilpotent
\begin{align}
 {\bf G}^2=0\,,
\end{align}
which in turns implies that the gauge symmetry \eqref{boldgauge} is infinitely reducible, just as in the boundaryless case.

\subsubsection{BV master action}
Consider the space of supervectors at degree $d$
\begin{align}
{\bf H}^{(d)}=\left(\begin{matrix} H^{(d)} \\ H^{(d-1)} \end{matrix}\right)\,.
\end{align}
Together, the $\bf{Q}$, $\bf{G}$ and $(-\bf{G}^\star)$ operators act as a nilpotent map between these spaces 
\begin{align}
\cdots\overset{\bf{G}}{\longrightarrow}{\bf H}^{(-2)}\overset{\bf{G}}{\longrightarrow}{\bf H}^{(-1)}\overset{\bf{G}}{\longrightarrow}{\bf H}^{(0)}\overset{\bf{Q}}{\longrightarrow}{\bf H}^{(1)}\overset{-\bf{G}^\star}{\longrightarrow}{\bf H}^{(2)}\overset{-\bf{G}^\star}{\longrightarrow}{\bf H}^{(3)}\overset{-\bf{G}^\star}{\longrightarrow}\cdots
\end{align}
This sequence of maps is nilpotent because of the properties
\begin{subequations}
\begin{align}
{\bf{Q}}^\star&=-{\bf{Q}}\,,\\
{\bf{G}}^2&=0\,,\\
{\bf{Q}}{\bf{G}}&=0\,,\\
-{\bf{G}}^\star{\bf{Q}}&=0\,,\\
(-{\bf{G}}^\star)^2&=0\,.
\end{align}
\end{subequations}
This allows us to define the nilpotent super BPZ-odd operator
\begin{subequations}
\begin{align}
{\bf Q}_{BV}&\equiv{\bf{G}}{\bf{P}}_{d\leq -1}+{\bf{Q}}{\bf{P}}_{d=0}-{\bf{G}}^\star{\bf{P}}_{d\geq1}\,,\\
{\bf Q}_{BV}^2&=0\,,\label{BV-nil}\\
{\bf Q}_{BV}^\star&=-{\bf Q}_{BV}\,\label{BV-cycl},
\end{align}
\end{subequations}
where ${\bf{P}}_d$ projects on the (super)-degree $d$ subspace of the full Hilbert space.\footnote{Notice that ${\bf{P}}^\star_{d}={\bf{P}}_{1-d}$ and ${\bf Q}_{BV}{\bf{P}}_d={\bf{P}}_{d+1}{\bf Q}_{BV} $.}

The construction of the BV master action is now standard. The BV string field is a state in ${\bf H}=\oplus_{d\in \mathbb{Z}}{\bf H}^{(d)}$ 
\begin{align}
\boldsymbol{\Psi}_{BV}= \left(\begin{matrix}\Psi_{BV}\\\chi_{BV} \end{matrix}\right)=\sum_{d\in \mathbb{Z}}\left(\begin{matrix} \sum_{s_d} \psi_{s_d}^{(-d)} \ket{{s}_d} \\  \sum_{s_{d-1}} \chi_{s_{d-1}}^{(-d)}\ket{{s}_{d-1}} \end{matrix}\right),
\end{align}
where $\ket{s_d}$ is a basis of  $H^{(d)}$ and the spacetime  fields $\psi_{s_d}^{(-d)}$ and  $\chi_{s_{d-1}}^{(-d)}$ are the (bulk and  boundary) BV fields with BV grading given by $(-d)$. BV spacetime fields with positive or vanishing BV grading are considered as fields while the others are considered as anti-fields.  The grassmanality of the spacetime BV fields is given by the BV grading modulo 2. Notice importantly that despite the fact that the basis vectors  $\ket{{ s}_d}$ and  $\ket{{ s}_{d-1}}$ have opposite grassmanality, the space-time BV fields $\psi_{s_d}^{(-d)}$ and  $\chi_{s_{d-1}}^{(-d)}$  have the same grassmanality induced by the overall BV degree. In particular  the total BV string field $\boldsymbol{\Psi}_{BV}$ is a supervector of total degree 0, by summing the super-degree of the basis vectors and the BV degree of the spacetime fields.
With this preparation, the classical BV master action is given by the standard expression
\begin{align}
S_{BV}=\frac12 \w\left(\boldsymbol{\Psi}_{BV}^T, {\bf Q}_{BV}\boldsymbol{\Psi}_{BV}\right)\,,
\end{align}
which correctly reduces to the classical gauge-invariant action \eqref{Stotbf} when the anti-fields are set to zero.
The classical master equation is then easily shown to be satisfied as
\begin{align}
{\Big (}S_{BV},S_{BV}{\Big)}_{\!BV}&=\omega\left(\left(\frac{\delta S_{BV}}{\delta \boldsymbol{\Psi}_{BV}}\right)^T, \frac{\delta S_{BV}}{\delta \boldsymbol{\Psi}_{BV}} \right)=\omega(({\bf Q}_{BV}\boldsymbol{\Psi}_{BV})^T,{\bf Q}_{BV}\boldsymbol{\Psi}_{BV})\nonumber\\
&=\omega(\boldsymbol{\Psi}_{BV}^T,{\bf Q}_{BV}^\star{\bf Q}_{BV}\boldsymbol{\Psi}_{BV})= -\omega(\boldsymbol{\Psi}_{BV}^T,{\bf Q}_{BV}^2\boldsymbol{\Psi}_{BV})=0\,,
\end{align}
where we have used the super-cyclicity \eqref{BV-cycl} and nilpotency \eqref{BV-nil} of ${\bf Q}_{BV}$. 

As a final comment on the construction we have just presented, it is noticeable that the presence of a boundary in target space is only visible in the physical sector at BV degree 0 (i.e. in the gauge-invariant action). At different BV grading, there is no notion of boundary anymore and the corresponding BV fields are defined on $\mathbb{R}^D$. The reason for this is that the `kinetic' terms of the non-physical BV fields are directly inherited from the structure of the (reducible) gauge transformations (rather than the equations of motion) and, by assumption of our construction, such gauge transformations are essentially the same as in the boundary-less case. To be more precise, the variation of the BV master action with respect to the anti-fields of a given BV degree should give the gauge-transformations \eqref{boldgauge} at the corresponding degree of reducibility (with the gauge parameter replaced by the corresponding ghost field) and these gauge transformations do not know about the boundary. It would be interesting to investigate whether $S_\mathrm{tot}$ could be recast in the BV-BFV framework of \cite{Cattaneo:2012qu} which is designed to deal with gauge theories on manifolds with boundaries. 

\section{Examples}\label{sec:examples}
In this section we will study specific examples of open and closed string backgrounds. For the sake of simplicity, we will specialize on the case of a flat boundary, taking our target space to be the half-space
\begin{align}\label{eq:halfspace}
M=\mathbb{R}_+\times \mathbb{R}^{D-1}\,.
\end{align}
This is a simple enough choice where we can explicitly see our construction at work in a non-trivial way and at the same time allows for quite manageable examples of bulk+boundary actions.
Let $z$ be the global coordinate on $\mathbb{R}_+$ so that the boundary is at $z=0$.
The normal vector to the boundary (pointing outside $M$) is constant and is given by
\begin{align}
n^\mu=\big(-1,\mathbf{0}\big)\,.
\end{align}
The normal derivative is therefore identified with the $z$-derivative
\begin{align}
\del_\perp \equiv- n^\mu\del_\mu=\del_z\,.
\end{align}
The string field in position space $\Psi(x)$ is now  understood as
\begin{align}
\Psi(x)=\Psi(z,y)\,,
\end{align}
where we will often keep the dependence on the boundary coordinates $y^a$ implicit.
In this representation the  BRST operator can be written as
\begin{align}
Q=-c\del_z^2-\Omega\del_z+\tilde Q\,.\label{Qderz}
\end{align}
In particular, for the open string, we define
\begin{subequations}
\begin{align}
c&=\alpha' c_0\,, \\
\Omega&=i\sqrt{2\alpha'}\sum_{n\neq 0}c_{-n} \alpha^z_n\,,
\end{align}
\end{subequations}
while for the closed string, we write
\begin{subequations}
\begin{align}
c&=\frac{\alpha'}{2} c_0^+\,, \\
\Omega&=i\sqrt{\frac{\alpha'}{2}}\sum_{n\neq 0}\left(c_{-n} \alpha^z_n+\bar c_{-n} \bar\alpha^z_n\right)\,,
\end{align}
\end{subequations}
where $\tilde Q$ is free of $z$ derivatives (but not of $y$ derivatives).
In this representation the algebraic relations which are responsible for the nilpotency of $Q$ are
\begin{subequations}
\label{eq:Qalg}
\begin{align}
c^2&=0\,,\\
[c,\tilde Q]-\Omega^2&=0\,,\\
\tilde  Q^2&=0\,,\\
[\Omega,c]&=0\,,\\
[\Omega, \tilde Q]&=0\,,
\end{align} 
\end{subequations}
where $[\cdot,\cdot]$ is the graded commutator. 
We can express the $\mathbb{R}^D$ symplectic form as
\begin{align}
\w(\psi_1,\psi_2)=\int_{\mathbb R}dz\,\w_z(\psi_1(z),\psi_2(z))\,,
\end{align}
where $\w_z(\cdot,\cdot)$ takes care of computing the BPZ inner product for all modes, including the center of mass along the boundary but excluding the center of mass in the $z$-direction, namely
\begin{align}
\w_z(\psi_1(z),\psi_2(z))\equiv\int_{\mathbb{R}^{D-1}}d^{D-1}y \,\w'(\psi_1(z,y),\psi_2(z,y))\,.
\end{align}
Placing the flat boundary at $z=0$ has the effect of changing the symplectic form as
\begin{align}
\w(\psi_1,\psi_2)\longrightarrow\w_\Theta(\psi_1,\psi_2)\equiv \w(\psi_1,\Theta\,\psi_2)=\int_{\mathbb R^+}dz\,\w_z(\psi_1(z),\psi_2(z))\,,
\end{align}
where the distribution-like operator $\Theta(z)$ is just the characteristic function of the positive real axis. 
%
We also have
\begin{align}
[\Theta,Q]=B=c\del_z\delta(z)+\delta(z) c\del_z+\delta(z)\Omega\equiv  \delta\Gamma^*+\Gamma\delta\,,\label{B-Gammaz}
\end{align}
 where 
\begin{subequations}
\begin{align}
\Gamma&= c\del_z-\gamma\,,\\
\Gamma^*&=c\del_z-\gamma^*\,,
\end{align}
\end{subequations}
and where $\gamma$ is a ghost-number one operator containing no $z$ derivatives which obeys
\begin{align}
\gamma+\gamma^*=-\Omega\label{gamma}\,.
\end{align}
For the sake of concreteness, in the explicit examples below, we will be making a particular choice $\gamma=\gamma^*$. In such a case we can write
\begin{align}
\Gamma_{\rm simple} ^*=c\,\p_z
+\frac{1}{2}\Omega 
\,.\label{Gsimple}
\end{align}
We will see that this choice of $\Gamma^*$ leads to structurally simple forms of boundary terms. This can be traced back to the observation that as soon as we make $\gamma^*$ BPZ-even, it drops out from the action, as we can clearly see from the factorized form \eqref{eq:StotGI}.

Considering now the gauge-invariant action \eqref{Stot}
\begin{align}
S_{\rm tot}(\Psi,\chi)=\frac12 \w(\Psi, (\Theta Q-\delta \Gamma^*)\Psi)+\w(\Psi,\delta\Gamma^* Q\chi)-\frac12\w\left(Q\chi,\delta\Gamma^*\,Q\chi\right),\label{Stotz}
\end{align}
it depends on the three boundary fields $(\chi_{-1}(y),\chi_{0}(y), \chi_{1}(y))$ which were defined by \eqref{eq:chis}. These can be equivalently thought of as arising from the Taylor expansion of $\chi(z)$ around $z=0$, namely
 \begin{align}\label{eq:Taylorexp}
 \chi(z)=\chi_{-1}+z\,\chi_{0}+\frac12 z^2\, \chi_{1}+O(z^3)\,.
 \end{align}
Analogously, one may expand the gauge parameter near the boundary as
\begin{align}
 \Lambda(z)=\Lambda(0)+z\,\Lambda'(0)+\frac12 z^2\, \Lambda''(0)+O(z^3)\,,
\end{align}
where the $'$ on $\Lambda(z)$ stands for $\del_z$, so that the $\Lambda$-gauge transformation of $\chi$ 
\begin{align}
\delta_\Lambda\chi=\Lambda\,,
\end{align}
translates into the gauge transformation
\begin{subequations}
\begin{align}
\delta_\Lambda\chi_{-1}&=\Lambda(0)\,,\\
\delta_\Lambda\chi_{0}&=\Lambda'(0)\label{chi-gauge}\,,\\
\delta_\Lambda\chi_{1}&=\Lambda''(0)\,,
\end{align}
\end{subequations}
of the boundary string fields $\chi_k(y)$.

\subsection{Massless open string}\label{subsec:massless_open}

The free action describing the open-string excitations at the (massless) level 0 can be generated from \eqref{Stot} by substituting 
\begin{subequations}
\begin{align}
    \Psi(x) &= A_\mu(x)\alpha_{-1}^\mu c_1|0\rangle -i\sqrt{\frac{\alpha^\prime}{2}}c_0 B(x)|0\rangle\,,\\
    \chi_k(y) &= \frac{i}{\sqrt{2\alpha^\prime}} \chi^{(k)}(y)|0\rangle\,.
\end{align}
\end{subequations}
for the bulk and boundary string fields. The corresponding gauge-parameter string field giving rise to the gauge transformations of the component fields $A_\mu, B$ and $\chi_k$ then reads
\begin{align}
    \Lambda(x) &= \frac{i}{\sqrt{2\alpha^\prime}} \lambda(x)|0\rangle\,.\label{eq:Lambda_lev0}
\end{align}
No non-trivial gauge-for-gauge parameter $\Upsilon$ arises at this level. Substituting the gauge-parameter $\Lambda(x)$ given by \eqref{eq:Lambda_lev0} into the general form \eqref{eq:gauge_both} of the gauge transformation of $\Psi$ and $\chi$, one finds
\begin{subequations}
    \begin{align}
        \delta_g A_\mu(x) &=\p_\mu \lambda(x)\,,\\
        \delta_g B(x) &= \p^2 \lambda(x)\,,\\
        \delta_g \chi_k(y) &= \p_z^{k+1}\lambda(x)|_{z=0}\,.
    \end{align}
\end{subequations}
At the same time, the consistent gauge-invariant action \eqref{Stot} turns into
\begin{align}
    S(A_\mu,B,\chi^{(k)})=\frac{\alpha^\prime}{2}\int d^d x \,\Big(\tfrac{1}{2}F_{\mu\nu}F^{\mu\nu}+\mathcal{K}^2\Big)+\frac{\alpha^\prime}{2}\int d^{d-1} y\, \mathcal{R}^{(0)}\mathcal{R}^{(1)}\,,\label{eq:So_lev0}
\end{align}
where the dependence on auxiliary fields (both in the bulk and on the boundary) was packaged into the gauge-invariant combinations
\begin{subequations}
\begin{align}
    \mathcal{K}&=B-\p_\mu A^\mu\,,\\
    \mathcal{R}^{(0)}&=\chi^{(0)}-A_z\,,\\
    \mathcal{R}^{(1)}&=\chi^{(1)}-\p^a\p_a\chi^{(-1)}+2\p_a A^a-B\,.
\end{align}
\end{subequations}
Noting that $\mathcal{K}$ can be set to zero by integrating out $B$ and, similarly, recognizing $\mathcal{R}^{(0)}$ as the equation of motion of the auxiliary field $\chi^{(1)}$, we conclude that \eqref{eq:So_lev0} is equivalent to the Maxwell action for the $U(1)$ gauge field $A_\mu$.

\subsection{First massive level of open string}\label{subsec:first_massive}

To test that the free action \eqref{Stot} is capable of consistently describing also the massive string excitations, let us expand it in the component fields arising at level 1. In particular, we should obtain an action equivalent to the one already presented in Section 5 of \cite{paper1}. To this end, we consider the following truncations of the bulk and boundary string fields
\begin{subequations}
\label{eq:SF_lev1}
    \begin{align}
    \Psi(x)&=\frac{1}{2}c_1 \varphi_{\mu\nu}(x)\alpha_{-1}^\mu\alpha_{-1}^\nu|0\rangle - ic_0\sqrt{\frac{\alpha^\prime}{2}}C_\mu(x)\alpha_{-1}^\mu|0\rangle+\frac{1}{2}c_{-1}D(x)|0\rangle+\nonumber\\
    &\hspace{5cm}+\frac{1}{2}ic_1 a_\mu(x)\alpha_{-2}^\mu|0\rangle +c_0b_{-2}c_1 \sqrt{\frac{\alpha^\prime}{2}}a(x)|0\rangle\,,\label{eq:Psi_lev1}\\
        \chi_k(y)&=\frac{i}{\sqrt{2\alpha^\prime}}\chi_\mu^{(k)}(y)\alpha_{-1}^\mu|0\rangle -\frac{1}{\sqrt{2\alpha^\prime}} \chi^{(k)}(y)b_{-2}c_1|0\rangle\,,
    \end{align}
\end{subequations}
together with the gauge parameter
\begin{align}
    \Lambda(x) = \frac{i}{\sqrt{2\alpha^\prime}}\xi_\mu(x)\alpha_{-1}^\mu|0\rangle -\frac{1}{\sqrt{2\alpha^\prime}} \xi(x)b_{-2}c_1|0\rangle\,.\label{eq:Lambda_lev1}
\end{align}
In $d=26$ dimensions, an action for the massive spin-2 triplet $(\varphi_{\mu\nu},C_\mu,D)$ of the target fields should be invariant under the gauge transformation
\begin{subequations}
    \begin{align}
        \delta_g\varphi_{\mu\nu}(x) &=\p_\mu\xi_\nu +\p_\nu \xi_\mu - \tfrac{1}{2}\eta_{\mu\nu}\sqrt{\tfrac{2}{\alpha^\prime}}\xi \,,\\[1mm]
        \delta_g C_{\mu}(x) &=\big(\p^2-\tfrac{1}{\alpha^\prime}\big)\xi_\mu (x)\,,\\
        \delta_g  D(x) &=2\p_\mu \xi^\mu - 3\sqrt{\tfrac{2}{\alpha^\prime}}\xi\,,
    \end{align}
\end{subequations}
which is indeed generated upon substituting \eqref{eq:Lambda_lev1} into \eqref{gauge1}. This has to be supplemented by the corresponding gauge transformation
\begin{subequations}
    \begin{align}
        \delta_g\chi_\mu^{(k)}(y) &= \p^{k+1}\xi_\mu(x)|_{z=0}\,,\\
        \delta_g\chi^{(k)}(y) &= \p^{k+1}\xi(x)|_{z=0}\,,
    \end{align}
\end{subequations}
of the boundary fields $\chi_\mu^{(k)}(y)$ and $\chi^{(k)}(y)$. After substituting the truncating string fields \eqref{eq:SF_lev1} into \eqref{Stot}, one obtains the gauge-invariant action
\begin{align}
&S(\varphi_{\mu\nu},C_\mu,D,a_\mu,a,\chi_\mu^{(k)},\chi^{(k)})=\nonumber\\
&\hspace{1cm}
= \frac{\alpha^\prime}{2}\int_{M}d^{26}x\, \bigg\{\tfrac{1}{2}\p_\rho\varphi_{\mu\nu}\p^\rho\varphi^{\mu\nu}+\tfrac{1}{2}\tfrac{1}{\alpha^\prime}\varphi^{\mu\nu}\varphi_{\mu\nu}-\tfrac{1}{2}\p_\mu D\p^\mu D-\tfrac{1}{2}\tfrac{1}{\alpha^\prime} D^2+\nonumber\\[-1mm]
&\hspace{8.3cm}+\p^\mu \varphi_{\mu\nu}\p^\nu D-\p_\nu\varphi_{\mu\rho}\p^\mu\varphi^{\nu\rho}+\nonumber\\[0.4mm]
&\hspace{4cm}+\tfrac{1}{4}(\p^\mu a^\nu-\p^\nu a^\mu)(\p_\mu a_\nu-\p_\nu a_\mu)+\nonumber\\[1.4mm]
&\hspace{4cm}+\sqrt{\tfrac{1}{2\alpha^\prime}}\Big[2\eta^{\rho\sigma}(\p_\rho \varphi_{\mu\sigma}-\p_\mu \varphi_{\rho\sigma})-\tfrac{5}{2}\p_\mu(D-\eta_{\rho\sigma}\varphi^{\rho\sigma})\Big]a^\mu+\nonumber\\
&\hspace{4cm}
-\tfrac{1}{16}\tfrac{1}{\alpha^\prime}(5D-\eta^{\mu\nu}\varphi_{\mu\nu})(D-\eta^{\mu\nu}\varphi_{\mu\nu})+\mathcal{K}_\nu\mathcal{K}^\nu+2\mathcal{L}^2\bigg\}\nonumber\\
    &\hspace{0.4cm}+\frac{\alpha^\prime}{2}\int_{\partial M}d^{25}y\, \bigg\{\varphi_{za}\Big(2\p^a\varphi_{zz}-\p^aD-2a^a\sqrt{\tfrac{1}{2\alpha^\prime}}\Big)+\nonumber\\
    &\hspace{4cm}-2\chi_z^{(-1)}\Big[\p^a\p_a(D-\varphi_{zz})-\p^a\p^b\varphi_{ab}+\nonumber\\
    &\hspace{5.7cm}+\p_a a^a \sqrt{\tfrac{2}{\alpha^\prime}}+\tfrac{1}{4}\tfrac{1}{\alpha^\prime}\eta_{ab}\varphi^{ab}-\tfrac{5}{4}\tfrac{1}{\alpha^\prime}D+\tfrac{5}{4}\tfrac{1}{\alpha^\prime}\varphi_{zz}\Big]+\nonumber\\[0.4mm]    &\hspace{4cm}+\mathcal{R}_a^{(0)}\mathcal{R}^{(1),a}+ \mathcal{R}_z^{(0)}\mathcal{R}_z^{(1)} +2\mathcal{R}^{(0)}\mathcal{R}^{(1)}\bigg\}\,,\label{eq:massive_lev1}
\end{align}
where, for the sake of transparency, we have again chosen to package all auxiliary fields in terms of the gauge-invariant combinations
\begin{subequations}
\begin{align}
    \mathcal{K}_\nu&=C_\nu - \p^\mu\varphi_{\mu\nu}+\tfrac{1}{2}\p_\nu D+\sqrt{\tfrac{1}{2\alpha^\prime}}a_\nu\,,\\
    \mathcal{L}&=
    a-\tfrac{1}{2}\p_\mu a^\mu -\tfrac{1}{4}\eta^{\mu\nu}\varphi_{\mu\nu}\sqrt{\tfrac{1}{2\alpha^\prime}}+\tfrac{3}{4}D\sqrt{\tfrac{1}{2\alpha^\prime}}\,,\\[0.7mm]
    \mathcal{R}_a^{(0)}&=\chi_a^{(0)}+\p_a \chi_z^{(-1)}-e_{za}\,,\\[0.4mm]    \mathcal{R}_a^{(1)}&=\chi^{(1)}_a+2\p_a\chi_z^{(0)}+2\p^b e_{ab}+\tfrac{1}{\alpha^\prime}\chi^{(-1)}_a-\sqrt{\tfrac{2}{\alpha^\prime}}a_a-e_a-\p_a e -\p_b\p^b \chi^{(-1)}_a\,,\\
    \mathcal{R}_z^{(0)}&=\chi_z^{(0)}+\tfrac{1}{2}e-e_{zz}-\p^a\chi_a^{(-1)}+\sqrt{\tfrac{2}{\alpha^\prime}}\chi^{(-1)}\,,\\[1mm]
    \mathcal{R}_z^{(1)}&=\chi_z^{(1)}+\p^a\p_a \chi_z^{(-1)}-\tfrac{1}{\alpha^\prime}\chi_z^{(-1)}-e_z\,,\\
    \mathcal{R}^{(0)}&=\chi^{(0)}-\tfrac{1}{2}a_z+\tfrac{1}{2}\sqrt{\tfrac{2}{\alpha^\prime}}\chi_z^{(-1)}\,,\\[-0.7mm]
    \mathcal{R}^{(1)}&=\chi^{(1)}-\p^a\p_a\chi^{(-1)}+\tfrac{1}{\alpha^\prime}\chi^{(-1)}+\p^a a_a-a+\tfrac{1}{4}\sqrt{\tfrac{2}{\alpha^\prime}}(e_{zz}+\eta_{ab}e^{ab}-3e+4\chi_z^{(0)}) \,.
\end{align}
\end{subequations}
Noticing that $\mathcal{K}_\nu$, $\mathcal{K}$, $\mathcal{R}_a^{(0)}$, $\mathcal{R}_z^{(0)}$ and $\mathcal{R}^{(0)}$ can all be set to zero upon integrating out the auxiliary fields $C_\nu$, $a$, $\chi_a^{(1)}$, $\chi_z^{(1)}$ and $\chi^{(1)}$, respectively, we conclude that we have indeed recovered the action reported in \cite{paper1}.

\subsection{Massless closed string}\label{subsec:massless_closed}
In this section we will describe the construction of the gauge-invariant action for the massless sector of closed-string. The level-0 component of the bulk closed string field is 
\begin{align}\label{eq:level0bulkstringfield}
\Psi(x)&=-\frac{1}{2}e_{\mu\nu}(\hat{x}) \alpha_{-1}^\mu \bar{\alpha}_{-1}^\nu c_1\bar{c}_1|0\rangle+e(\hat{x})c_1c_{-1}|0\rangle+\bar{e}(\hat{x})\bar{c}_1\bar{c}_{-1}|0\rangle+\nonumber  \\&
    \hspace{4cm}+i\sqrt{\frac{\alpha'}{2}} c_0^+ \big( e_\mu(\hat{x}) c_1\alpha_{-1}^\mu+\bar{e}_\mu(\hat{x}) \bar{c}_1\bar{\alpha}_{-1}^\mu\big)|0\rangle\,.   \end{align}
and the boundary string field is the most generic level-matched ghost number $1$ string field at level zero 
\begin{align}\label{eq:level0bundarystringfield}
    \chi_{k}(y)= \frac{i}{\sqrt{2\alpha'}}\big(\chi^{(k)}_{\mu}(y)c_1\alpha^{\mu}_{-1}-\bar{\chi}_{\mu}^{(k)}(y)\bar{c}_{1}\bar{\alpha}^{\mu}_{-1}\big)\ket{0}-\chi^{(k)}(y)c_{0}^{+}\ket{0}\,,
\end{align}
First, we shall focus on the gauge transformations. To do so, we introduce the string field gauge parameter at ghost number $1$ 
\begin{align}\label{eq:level0gaugestringfield}
    \Lambda= \frac{i}{\sqrt{2\alpha'}}\big(\xi_{\mu}(x)c_1\alpha^{\mu}_{-1}-\bar{\xi}_{\mu}(x)\bar{c}_{1}\bar{\alpha}^{\mu}_{-1}\big)\ket{0}-\xi(x)c_{0}^{+}\ket{0}\,,
\end{align}
and the string field gauge parameter at ghost number $0$
\begin{align}\label{eq:level0gaugeforgaugestringfield}
    \Upsilon= \frac{1}{\alpha'}u(x)\ket{0}\,,
\end{align}
which implements the gauge for gauge transformation $\delta_{g}\Lambda=Q\Upsilon$. These become relevant when we consider the transformations of boundary modes \eqref{gauge2}.  
At the level of target-space fields, the bulk gauge transformations $\delta_{g}\Psi=Q\Lambda$ read
\begin{subequations}
\begin{align}
    \delta_{g} e_{\mu\nu}(x)&=\partial_{\mu} \bar{\xi}_{\nu}(x)+\partial_{\nu}\xi_{\mu}(x)\,,\\
    \delta_{g} e_{\mu}(x)&=-\tfrac{1}{2}\partial^2\xi_\mu(x)-\partial_\mu\xi(x)\,,\\
    \delta_{g}\bar{e}_{\mu}(x)&=+\tfrac{1}{2}\partial^2\bar{\xi}_\mu(x)-\partial_\mu\xi(x)\,,\\
    \delta_{g}e(x)&= -\tfrac{1}{2}\partial_\mu \xi^\mu(x)-\xi(x)\,,\\
    \delta_{g}\bar{e}(x)&= +\tfrac{1}{2}\partial_\mu \bar{\xi}^\mu(x)-\xi(x)\,,
\end{align}
\end{subequations}
whereas the transformation laws for the boundary fields \eqref{gauge2} become
\begin{subequations}
\label{eq:gauge_closed_bndy}
\begin{align}
 \delta_{g}\chi_{\mu}(x)&= \xi_{\mu}(x)-\p_\mu u(x)\,,\\
      \delta_{g}\bar{\chi}_{\mu}(x)&= \bar{\xi}_{\mu}(x)+\p_\mu u(x)\,,\\
      \delta_{g}\chi(x)&=\xi(x)+\tfrac{1}{2}\p^2 u(x)\,.
\end{align}
\end{subequations}
For the sake of brevity, we wrote the gauge transformations for the boundary fields as if they were bulk fields, meaning that both the left-hand side and the right-hand side must be expanded according to \eqref{eq:Taylorexp}, thus obtaining the gauge transformation for each Taylor coefficient. For instance the third line of \eqref{eq:gauge_closed_bndy} yields the following boundary  gauge transformations 
\begin{align}
       \delta_g\chi_{{k}}(y)&=\partial^{k+1}_{z}\xi(x)|_{z=0}+\frac{1}{2}\p^{k+3}_{z}u(x)|_{z=0}+\frac{1}{2}\p^{k+1}_z\p_a^2  u(x)|_{z=0}\,,
\end{align}
for $k=-1,0,1$.

We now turn to the computation of the gauge-invariant action, obtained by substituting \eqref{eq:level0bulkstringfield} and \eqref{eq:level0bundarystringfield} into \eqref{Stot}. The resulting expression is
\begin{align}
\label{eq:closedlevel0action}
&S(e_{\mu\nu},e_\mu,\bar{e}_\mu,e,\bar{e};\chi^{(k)}, \chi^{(k)}_{\mu}, \bar{\chi}^{(k)}_{\mu})=\nonumber\\[1mm]
&\hspace{1cm}= \frac{\alpha'}{16}\int_{M}d^{d}x\, \Big[ -\tfrac{1}{4}\partial_{\rho}e_{\mu\nu}\partial^{\rho}e^{\mu\nu}+\tfrac{1}{4}\partial_{\rho}e_{\mu\nu}\partial^{\nu}e^{\mu \rho}+\tfrac{1}{4}\partial_{\rho}e_{\nu\mu}\partial^{\nu}e^{\rho\mu}\nonumber\\&\hspace{4.4cm}+\big(\partial_{\mu}(e-\bar{e})\big)^2-\tfrac{1}{2}\partial^\nu(e_{\nu\mu}+e_{\mu\nu})\partial^{\mu}(\bar{e}-e)-\mathcal{K}_\mu\mathcal{K}^\mu-\bar{\mathcal{K}}_\mu\bar{\mathcal{K}}^\mu\Big]\nonumber\\
    &\hspace{1cm}+ \frac{\alpha'}{32}\int_{\partial M}d^{d-1}y\, \Big[(e_{az}+e_{za})\partial^{a}(\bar{e}-e-e_{zz})+\nonumber\\
    &\hspace{5cm}-\partial_{a}\big(\bar{\chi}_{z}^{(-1)}+\chi_{z}^{(-1)}\big)\partial_b\big((2\bar{e}-2e-e_{zz})\eta^{ab}-e^{ab}\big)+\nonumber\\[1.4mm]
    &\hspace{5cm}-\mathcal{R}_a^{(0)}\mathcal{R}^{(1),a}-\bar{\mathcal{R}}_a^{(0)}\bar{\mathcal{R}}^{(1),a}-\mathcal{R}_z^{(0)}\mathcal{R}_z^{(1)}-\bar{\mathcal{R}}_z^{(0)}\bar{\mathcal{R}}_z^{(1)}\Big]
    \,,
\end{align}
where all dependence on the bulk and boundary auxiliary fields is captured by the gauge-invariant combinations
\begin{subequations}
    \begin{align}
        \mathcal{K}_\mu &=e_{\mu}+\tfrac{1}{2}\partial^{\nu}e_{\mu\nu}-\partial_{\mu}\bar{e}\,,\\
        \bar{\mathcal{K}}_\mu &=\bar{e}_{\mu}-\tfrac{1}{2}\partial^{\nu}e_{\nu\mu}-\partial_{\mu}e\,,\\
        \mathcal{R}_a^{(0)}&=\chi_a^{(0)}+\p_a\chi_z^{(-1)}-e_{az}\,,\\
        \bar{\mathcal{R}}_a^{(0)}&=\bar{\chi}_a^{(0)}+\p_a\chi_z^{(-1)}-e_{za}\,,\\
        \mathcal{R}_a^{(1)}&=\tfrac{1}{2}\chi^{(1)a}+\p^a\bar{\chi}_z^{(0)}-\p^a\chi^{(-1)}-\tfrac{1}{2}\p^2_b \chi^{(-1)a}+e^a-2\p ^a\bar{e}+\p_b e^{ab}\,,\\
        \bar{\mathcal{R}}_a^{(1)}&=\tfrac{1}{2}\bar{\chi}^{(1)a}+\p^a\chi_z^{(0)}+\p^a\chi^{(-1)}-\tfrac{1}{2}\p^2_b \bar{\chi}^{(-1)a}-\bar{e}^a+2\p ^a e+\p_b e^{ba}\,,\\
        \mathcal{R}_z^{(0)}&=\tfrac{1}{2}\chi_z^{(0)}+\chi^{(-1)}-\tfrac{1}{2}\p^a \bar{\chi}_{a}^{(-1)}+\bar{e}-\tfrac{1}{2}e_{zz}\,,\\
        \bar{\mathcal{R}}_z^{(0)}&=\tfrac{1}{2}\bar{\chi}_z^{(0)}-\chi^{(-1)}-\tfrac{1}{2}\p^a \chi_{a}^{(-1)}-e-\tfrac{1}{2}e_{zz}\,,\\
        \mathcal{R}_z^{(1)}&= \chi_z^{(1)}+2\chi^{(0)}+\p^2_a \chi_z^{(-1)}+2 e_z\,,\\
        \bar{\mathcal{R}}_z^{(1)}&= \bar{\chi}_z^{(1)}-2\chi^{(0)}+\p^2_a \bar{\chi}_z^{(-1)}-2\bar{e}_z\,.
    \end{align}
\end{subequations}
In particular, both $\mathcal{K}_\mu$ and $\bar{\mathcal{K}}_\mu$ go to zero when we integrate out classically the two bulk auxiliary fields $e_{\mu}$ and $\bar{e}_{\mu}$. Similarly, the combinations $\mathcal{R}_a^{(0)}$, $\bar{\mathcal{R}}_a^{(0)}$, $\mathcal{R}_z^{(0)}$ and $\bar{\mathcal{R}}_z^{(0)}$ can be set to zero by solving the equations of motion for the boundary auxiliary fields $\chi^{(1)}_{a}$, $\bar{\chi}^{(1)}_{a}$, $\chi^{(1)}_{z}$ and $\bar{\chi}^{(1)}_{z}$.  Therefore, after having integrated out the auxiliary fields, the only contribution from the boundary modes is due to $\chi_z^{(-1)}$ and $\bar{\chi}_z^{(-1)}$ which only appear through the combination
\begin{align}\label{eq:l}
    l(y)=\bar{\chi}_{z}^{(-1)}(y)+\chi_{z}^{(-1)}(y)\,,
\end{align}
which, according to \eqref{eq:gauge_closed_bndy}, transforms as
\begin{align}    \delta_{g}l(y)=\big( \xi_{z}
(x)+\bar{\xi}_{z}(x)\big)|_{z=0}\,.
\end{align}
Finally, to make contact with \cite{paper1}, we split the rank-2  tensor field $e_{\mu \nu} $ in its symmetric and anti-symmetric part 
\begin{align}
       e_{\mu \nu}=b_{\mu \nu}+h_{\mu\nu}\,,\qquad\qquad b_{\mu\nu}=-b_{\nu\mu}\,,\qquad\qquad h_{\mu\nu}=h_{\nu\mu}\,,
\end{align}
introduce the $3$-form
\begin{align}
    H_{\mu\nu\rho}=\p_\mu b_{\nu\rho}+\p_{\nu}b_{\rho \mu}+\p_{\rho}b_{\mu \nu}\,,
\end{align}
and perform the field redefinition
   \begin{align}\label{eq:physicaldilaton}
    \phi=\frac{1}{4}(\tensor{h}{^\mu_\mu}-2\bar{e}+2e)\,.
\end{align}
Therefore, the action \eqref{eq:closedlevel0action} can be rewritten as 
\begin{align}\label{eq:level0closedactionfinal2}
    &S^{\ast}\left(h_{\mu \nu},b_{\mu \nu},\phi;l\right)=\nonumber\\
    &\hspace{0.4cm}=-\frac{\alpha^\prime}{16}\int_{M}d^d x\, \bigg(\tfrac{1}{12}H_{\mu\nu\rho}H^{\mu\nu\rho}\bigg)+\nonumber\\
    &\hspace{1cm}-\frac{\alpha^\prime}{16}\int_{M}d^d x\,\bigg(\tfrac{1}{4}\partial_\rho h_{\mu\nu}\partial^\rho h^{\mu\nu}-\tfrac{1}{2}\partial_{\rho}h_{\mu \nu}\partial^\nu h^{\mu\rho}+\tfrac{1}{4}\p^\mu\big(2h_{\mu \nu}-\eta_{\mu \nu}\tensor{h}{^{\rho}_{\rho}}\big)\p^\nu \tensor{h}{^{\lambda}_{\lambda}}\nonumber\\
    &\hspace{7cm}-2\p^\mu\big(h_{\mu\nu}-\eta_{\mu \nu}\tensor{h}{^{\rho}_{\rho}}\big)\p^{\nu}\phi-4\p_\mu\phi\p^\mu\phi\bigg)+\nonumber\\
    &\hspace{1cm}+\frac{\alpha^\prime}{16}\int_{\partial M}d^{d-1} y\, \bigg[\tfrac{1}{2}h_{az}\partial^{a}\big(\tensor{h}{^{b}_{b}}-h_{zz}-4\phi\big)+\tfrac{1}{2}\left(\left(\tensor{h}{^{c}_{c}}-4\phi\right)\eta^{ab}-h^{ab}\right)\partial_{a}\partial_{b}l\bigg]\,,
\end{align}
which is invariant under the gauge transformations
\begin{subequations}
    \begin{align}
        \delta_{\Lambda}b_{\mu\nu}(x)&=\partial_{\mu}\lambda_{\nu}(x)-\partial_{\nu}\lambda_{\mu}(x)\,,\\
        \delta_{\Lambda}h_{\mu\nu}(x)&=\partial_{\mu}\tau_{\nu}(x)+\partial_{\nu}\tau_{\mu}(x)\,,\label{eq:deltah}\\
        \delta_{\Lambda}\phi(x)&=0\,,\\
        \delta_{\Lambda}l(y)&=2\tau_{z}(y)\,,\label{eq:deltal}
    \end{align}
\end{subequations}
where we defined the the rotated gauge parameters as
\begin{subequations}\label{eq:rotatedgaugeparameter}
    \begin{align}
      \lambda_\mu&=\frac{1}{2}(\bar{\xi}_\mu-\xi_{\mu})\,,\\
      \tau_\mu&=\frac{1}{2}(\bar{\xi}_\mu+\xi_{\mu})\,.
    \end{align}
\end{subequations}
Looking at \eqref{eq:level0closedactionfinal2}, we recognize the action previously obtained in \cite{paper1}, which describes the dynamics of the graviton, Kalb–Ramond, and dilaton in flat half-space where the boundary is free to fluctuate. Indeed, as extensively discussed in \cite{paper1}, \eqref{eq:level0closedactionfinal2} can be obtained by linearizing the standard low-energy effective action for gravity with the Kalb–Ramond and the dilaton, including the Gibbons–Hawking–York term
\begin{align}
\label{eq:GRaction}
    S_{\rm st}= \int_{M}d^dx\, \sqrt{g}\,e^{-2\phi}\Big(R-\tfrac{1}{12}H^2+4 (\nabla \phi)^2\Big)+\int_{\p M}d^{d-1}y \,\sqrt{h}\, e^{-2\phi}\, 2K\,,
\end{align}
where $R$ is the Ricci scalar, $K$ is the trace of the extrinsic curvature, and the boundary is defined through the implicit equation
\begin{align}
    F(x)=z+\frac{1}{2}l(y)=0\,,\label{eq:bndy_fluct}
\end{align}
which implies that the boundary mode $l$ can be interpreted as the transverse displacement of the fluctuating flat boundary.

\subsection{Linear dilaton backgrounds}\label{subsec:lin_dilaton}
In this section, we are going to apply our construction in the case of the linear dilaton CFT, 
with the goal of constructing the gauge-invariant action for the level-$0$ open and closed string sector in the half-space. 
\begin{myblock}
Before doing that, however, it is important to carefully understand how the linear dilaton background arises in General Relativity, in the presence of a (fluctuating) boundary.
\subsubsection{Linear dilaton with boundary in General Relativity}

 From the GR perspective, the linear dilaton background arises as a solution to the equations of motion of the Einstein-Hilbert action with dilaton and a cosmological constant $\Lambda$
\begin{align}
    S_{\rm bulk}=\int_{M}d^dx\, \sqrt{g}\,e^{-2\Phi}\Big(R+4 (\nabla \Phi)^2+\Lambda\Big)\,.
\end{align}
This action is the $d$-dimensional extension of the two-dimensional Callan-Giddings-Harvey-Strominger (CGHS) model \cite{Callan:1992rs}. Without a boundary, it admits the following solution
\begin{subequations}
\label{eq:lindilsol}
\begin{align}
        g^{\ast}_{\mu \nu}&= \eta_{\mu \nu}\,,\\
        \Phi^{\ast}&=\frac{1}{2}V_\mu x^\mu\,,
\end{align}
\end{subequations}
namely the flat metric and a linear profile for the dilaton, whose slope $V_\mu$ is related to the cosmological constant as
\begin{align}
    \Lambda=\eta^{\mu \nu}V_\mu V_\nu\,.
\end{align}
Interestingly, if we now specify the background  to be the flat half-space \eqref{eq:halfspace} with a flat fluctuating boundary, it turns out that the usual GHY term is not sufficient anymore to ensure a well-posed variational principle, as we will see shortly. To address this problem, we introduce a boundary cosmological constant $\Lambda_{\rm b}$, leading to the action
\begin{align}\label{eq:CGHSandboundary}
    S_{\rm ld}=\int_{M}d^dx\, \sqrt{g}\,e^{-2\Phi}\Big(R+4 (\nabla \Phi)^2+\Lambda\Big)+\int_{\p M}d^{d-1}y \,\sqrt{h}\, e^{-2\Phi}\, 2(K+\Lambda_{\rm b})\,.
\end{align}
In order for the solution \eqref{eq:lindilsol} to continue to extremize the action without the necessity to impose any restrictions on fluctuations at $\p M$, it turns out that we must impose the following relation between the dilaton slope and the boundary cosmological constant
\begin{align}\label{eq:condlambdab}
    \Lambda_{\rm b}=\eta^{\mu z}V_{\mu}\,.
\end{align}
This becomes clear by studying the fluctuations around the solution \eqref{eq:lindilsol}
\begin{subequations}
\begin{align}
         g_{\mu\nu}&=\eta_{\mu \nu}+h_{\mu\nu}\,,\\
         \Phi&=\frac{1}{2}V_\mu x^\mu+\phi\,,
\end{align}
\end{subequations}
together with the boundary fluctuation $l$ defined through \eqref{eq:bndy_fluct}.
Using standard GR manipulations (see appendix A of \cite{paper1} for details) and after a lengthy but straightforward computation, we find that the action  receives the following contributions, up to second order in fluctuations. First, a constant term due to the cosmological constants
   \begin{align}
       S^{(0)}_{\rm ld}= 2\Lambda\int_{M}d^dx\, e^{-V_\rho x^\rho}+2\Lambda_{\rm b}\int_{\partial M}d^{d-1}y\, e^{-V_a x^a}\,,
   \end{align}
 then a linear contribution, a potential boundary tadpole
   \begin{align}
       S^{(1)}_{\rm ld}=(\Lambda_{\rm b}-V_z)\int_{\partial M}d^{d-1}y\, e^{-V_a x^a} (h_a^a-4\phi+V_zl)\,.
   \end{align}
Remarkably this vanishes precisely when \eqref{eq:condlambdab} is satisfied, ensuring that \eqref{eq:lindilsol} extremizes the action without restricting the fluctuations $h_a^a$, $\phi$ and $l$.

Finally, one obtains a second-order contribution involving, in particular, the  term 
   \begin{align}
       S^{(2)}_{\rm ld}=\ldots-\frac{1}{2}(\Lambda_{\rm b}-V_z)\int_{\partial M}d^{d-1}y\, e^{-V_a x^a} l\,\partial_z(  h_a^a-4\phi)+\dots\,,
   \end{align}
  which,  for generic $V_z$ and $\Lambda_{\rm b}$, is plagued by transverse derivatives and thus makes the variational principle ill-defined.  But again, the relation \eqref{eq:condlambdab}  ensures the consistency of the expanded action.
After a series of other cancellations, the full second-order contribution to the linearized action turns out to read
\begin{align}\label{eq:CGHSsecond}
    &S_{\rm ld}^{(2)}=\nonumber\\
    &\hspace{0.2cm}=-\int_{M}d^d x\,e^{-V_\rho x^\rho}\bigg(\tfrac{1}{4}\partial_\rho h_{\mu\nu}\partial^\rho h^{\mu\nu}-\tfrac{1}{2}\partial_{\rho}h_{\mu \nu}\partial^\nu h^{\mu\rho}+\tfrac{1}{4}\big(2(\p^\mu-V^\mu)h_{\mu\nu}-\p_\nu \tensor{h}{^{\rho}_{\rho}}\big)\p^\nu \tensor{h}{^{\lambda}_{\lambda}}\nonumber\\
    &\hspace{7.2cm}-2\big((\p^\mu-V^\mu)h_{\mu\nu}-\p_{\nu}\tensor{h}{^{\lambda}_{\lambda}}\big)\p^\nu \phi -4\p_\mu\phi\p^\mu\phi\bigg)+\nonumber\\
    &\hspace{0.8cm}+\int_{\partial M}d^{d-1} y\,e^{-V_c y^c} \bigg[\tfrac{1}{2}h_{az}\partial^{a}\big(\tensor{h}{^{b}_{b}}-h_{zz}-4\phi\big)-\tfrac{1}{2}h^{ab}\p_a\p_b l-\tfrac{1}{2}\p^a l\p_a(\tensor{h}{^{b}_{b}}-4\phi)\nonumber\\
    &\hspace{10cm}-\tfrac{1}{4}V_z h_{zz}^{2}-\tfrac{1}{4}V_z \p_a l\p^a l\bigg]\,.    
\end{align}
This is a well-defined quadratic action with boundary and it is the expression that has to be compared with our quadratic SFT action defined on the linear dilaton background, with the understood inclusion of the Kalb–Ramond sector, which can be studied separately and is not affected by extra boundary subtleties.
\end{myblock}

\subsubsection{Linear dilaton with boundary in String Field Theory}
Let's start by briefly reviewing some basic aspects of the linear dilaton CFT. The main difference with respect to the free boson CFT is that the energy-momentum tensor now has an additional term proportional to the world-sheet second derivative of the scalar fields
\begin{align}\label{eq:Tld}
    T(z)=-\frac{1}{\alpha'}\eta_{\mu \nu}:\p X^\mu\p X^\nu:-i\sqrt{\frac{2}{\alpha'}}\beta_\mu\p^2X^\mu\,,
\end{align}
\begin{myblock}
  where $\beta_\mu$ is the dimensionless background charge, which we will later relate to $V_\mu$.
\end{myblock}
This gives rise to a cubic pole in the OPE with $\p X^{\mu}$, making this field no longer a primary. Indeed, it is possible to show that under the application of a generic conformal map $f(z)$, it obeys the following transformation rule
\begin{align}
    f\circ \p X^{\mu}(z)= f'(z)\p X^{\mu}(f(z))-i\sqrt{\frac{\alpha'}{2}}\beta^\mu \frac{f''(z)}{f'(z)}\,.
\end{align}
Consequently, this implies, among other things, that the matter oscillators behave differently under BPZ conjugation. In particular,
\begin{equation}
    (\alpha_{n}^{\mu})^\ast= \begin{cases}
        -\alpha_{-n}^\mu-2 \beta^\mu \delta_{n,0}& \hspace{2cm}\text{closed string}\,,\\
        (-)^{n+1}\alpha_{-n}^{\mu}-2\beta^{\mu}\delta_{n,0}& \hspace{2cm}\text{open string}\,.
    \end{cases}
\end{equation}
Therefore, from this point of view, the only difference with respect to the free boson CFT is related to the matter zero mode, a.k.a. the momentum
\begin{align}\label{eq:momentumBPZdual}
    (p_{\mu})^\ast=-p_\mu-iV_\mu\,,
\end{align}
where we used the dimensionful background charge 
\begin{align}
    V_\mu= \begin{cases}
        -2i\sqrt{\frac{2}{\alpha'}} \beta_\mu&\hspace{2cm}\text{closed string}\,,\\
        -i\sqrt{\frac{2}{\alpha'}}\beta_\mu& \hspace{2cm} \text{open string}\,,
    \end{cases}
\end{align}
in order to treat the open and closed string sectors on the same footing. Notice that the square of the background charge is fixed by the criticality condition of the theory 
\begin{align}
    c_{\beta}=d-12\beta^2=26\quad\longleftrightarrow\quad V^2= \begin{cases}
        -\frac{2(d-26)}{3\alpha'}&\hspace{2cm}\text{closed string}\,,\\
       -\frac{d-26}{3\alpha'}&\hspace{2cm}  \text{open string}\,.
    \end{cases}
\end{align}
In the position space representation, \eqref{eq:momentumBPZdual} becomes
\begin{align}\label{eq:delstarlineardilaton}
    (\p_\mu)^\ast = -\p_\mu+ V_\mu\,.
\end{align}
This property can be naturally implemented by defining the inner product as follows
\begin{align}\label{eq:lineardilatonsymplecticform}
    \omega_{V}(\Phi,\Psi)=\int_{\mathbb{R}^d}d^dx\,e^{-V_\mu x^\mu} \omega'(\Phi(x),\Psi(x))\,.
\end{align}
Exactly as in the free boson CFT, also in this framework the BRST charge can be decomposed in the position representation as
\begin{equation}
    Q_{V}=- c \p_\mu \p^\mu- \Omega_V^\mu \p_\mu+Q_{V}'\,,
\end{equation}
where  $\Omega^{\mu}_{V}$ and $Q'_{V}$ do not contain the position zero mode or its derivatives. These operators satisfy, by construction, the algebra \eqref{eq:Qalg} and obey the following BPZ properties
\begin{subequations}
    \begin{align}
        c^\ast&=-c,\\     (\Omega_{V}^{\mu})^\ast&=+\Omega_{V}^{\mu} + 2cV^\mu,\\
        (Q'_{V})^\ast&=-Q'_{V}+\Omega_V^{\mu}V_\mu+cV^2,
    \end{align}
\end{subequations}
which together with \eqref{eq:delstarlineardilaton}, guarantee that
\begin{align}
     (Q_{V})^\ast =-Q_{V}\,.
\end{align}
Note that in this context, the first-order differential operator $\Gamma_{V}^{\ast}$, which is decomposed according to  \eqref{eq:GammaDecomp}
\begin{align}
  \Gamma_{V}^{\ast }=c\p_z-\gamma_{V}^{\ast}\,.   
\end{align}
transforms under BPZ conjugation as 
\begin{align}
     \Gamma_{V}&=c\p_z -V_zc- \gamma_{V}\,.
\end{align}
Consequently, the defining relation \eqref{B-Gammaz} implies 
\begin{align}
    \gamma_{V}+\gamma_{V}^{\ast}=-\Omega_{V}-V_{z}c\,,\label{gammaV}
\end{align}
which is, as expected, a BPZ even combination. Therefore, analogous to \eqref{Gsimple} we can choose the `simple' form of $\gamma_{V}^{\ast}$
\begin{align}
    \gamma_{V}^{\ast}=\gamma_{V}=-\frac{1}{2}\left(\Omega_{V}+V_z c\right)\,,
\end{align}
leading to  
\begin{align}
    \Gamma_{V}^\ast=c\p_z+\frac{1}{2}\Omega_{V}+\frac{1}{2}V_z c\,.
\end{align}
This, unlike its counterpart in the free boson scenario, is not BPZ even. Nevertheless, also in this background, all the requirements for our construction of $S_\mathrm{tot}$ are satisfied. Therefore, we can compute a gauge-invariant action simply  using the symplectic form \eqref{eq:lineardilatonsymplecticform} and the operators $Q_{V}$ and $\Gamma_{V}$ in \eqref{Stot}.  

Let us now consider the explicit example of the level-0 open string sector. To ease manipulations, we can consistently truncate the BRST charge by identifying
\begin{subequations}
    \begin{align}
        c&=\alpha'c_{0}\,,\\
        \Omega_{V}^\mu
&=i\sqrt{2\alpha'}(c_1\alpha_{-1}^\mu+c_{-1}\alpha_1^\mu)    -\alpha' V^\mu c_0\,,\\
Q'_{V}&=-2b_0 c_{-1} c_1+i\sqrt{2\alpha'}V_{\mu}c_{-1}\alpha^{\mu}_{1}\,.
\end{align}
\end{subequations}
The field content we consider is the same as discussed in Section~\ref{subsec:massless_open} for both the bulk and boundary degrees of freedom. The first difference with respect to the case $V_\mu=0$ arises in the gauge transformations, which become
\begin{subequations}
    \begin{align}
        \delta_{g}A_{\mu}(x)&=\p_\mu \lambda(x)\,,\\
        \delta_{g}B(x)&=(\p_\mu-V_\mu)\p^\mu \lambda(x)\,,\\
        \delta_{g}\chi^{(k)}(y)&=\p^{k+1}_{z}\lambda(x)|_{z=0}\,.
    \end{align}
\end{subequations}
We now have all the ingredients to evaluate the gauge-invariant action \eqref{Stot}, which turns out to be
\begin{align}
S(A_\mu,B;\chi^{(k)})
 &=\frac{\alpha'}{2}\int_{M}d^{d}x\, e^{-V_\rho x^\rho}\Big(\tfrac{1}{2}F_{\mu\nu}F^{\mu\nu}+\mathcal{K}^2\Big)+\frac{\alpha'}{2}\int_{\p M}d^{d-1}y\, 
e^{-V_b y^b}\,\mathcal{R}^{(0)}\mathcal{R}^{(1)}\,,\label{eq:openlevel0actionld}
\end{align}
where we have denoted the gauge-invariant auxiliary structures
\begin{subequations}
    \begin{align}
        \mathcal{K}&=B-(\p^\mu-V^\mu)A_\mu\,,\\
        \mathcal{R}^{(0)}&=\chi^{(0)}-A_z\,,\\
        \mathcal{R}^{(1)}&=\chi^{(1)}-(\p_a-V_a)\p^a\chi^{(-1)}+2(\p_a-V_a)A^a-V_zA_z-B\,.
    \end{align}
\end{subequations}
As a sanity check, we can see that this action indeed reproduces \eqref{eq:So_lev0} for $V_\mu=0$.  
Furthermore, by solving the equations of motion for the auxiliary fields $B$ and $\chi^{(1)}$, the structures $\mathcal{K}$, $\mathcal{R}^{(0)}$, $\mathcal{R}^{(1)}$ can be removed to obtain 
\begin{align}\label{eq:masslessopenlineardilatonfinal}
    S^{\ast}(A_\mu)=\frac{\alpha'}{4}\int_{M}d^{d}x\, e^{-V_\rho x^\rho}F_{\mu\nu}F^{\mu\nu}\,.
\end{align}
The presence of a background charge is entirely encapsulated in the exponential prefactor. The action \eqref{eq:masslessopenlineardilatonfinal} reproduces the result appearing in \cite{Hellerman:2008wp} where, however, boundary terms were not present. This is parallel to what happens to the gauge-field in standard flat space-time.

The situation becomes somewhat more involved for the closed string action at level zero.  In this case, the operators that constitute the BRST charge can be consistently truncated to
\begin{subequations}\label{eq:closedQlineardilaton}
     \begin{align}
  c &= \tfrac{\alpha'}{2}c_0^+\,,\\
    \Omega_{V}^\mu &= i\sqrt{\tfrac{\alpha'}{2}}\big(c_1 \alpha_{-1}^\mu+c_{-1}\alpha_1^\mu+\bar{c}_1 \bar{\alpha}_{-1}^\mu+\bar{c}_{-1}\bar{\alpha}_1^\mu\big) -\tfrac{\alpha'}{2}c_{0}^{+}V^{\mu}\,,\\
    Q_{V}'&=-b_0^+ \big(c_{-1}c_{1}+\bar{c}_{-1}\bar{c}_1\big)+i\sqrt{\tfrac{\alpha'}{2}}V_\mu(c_{-1}\alpha_1^{\mu}+\bar{c}_{-1}\bar{\alpha}_{1}^{\mu})\,.
\end{align}
  \end{subequations}
In parallel with the definitions given in Section~\ref{subsec:massless_closed}, namely considering \eqref{eq:level0bulkstringfield} and \eqref{eq:level0bundarystringfield} as the bulk and boundary string fields, respectively, \eqref{eq:level0gaugestringfield} as the gauge string field parameter, and \eqref{eq:level0gaugeforgaugestringfield} as the gauge-for-gauge string field parameter, we obtain that the gauge transformations \eqref{gauge1} and \eqref{gauge2} correspond, at the level of the target space fields, to 
\begin{subequations}\label{eq:gaugetransflv0cld}
\begin{align}
    \delta_{g} e_{\mu\nu}(x)&=\partial_{\mu} \bar{\xi}_{\nu}(x)+\partial_{\nu}\xi_{\mu}(x)\,,\\
    \delta_{g} e_{\mu}(x)&=-\tfrac{1}{2}(\p^\nu-V^\nu)\partial_\nu\xi_\mu(x)-\partial_\mu\xi(x)\,,\\
    \delta_{g}\bar{e}_{\mu}(x)&=+\tfrac{1}{2}(\p^\nu-V^\nu)\partial_\nu\bar{\xi}_\mu(x)-\partial_\mu\xi(x)\,,\\
    \delta_{g}e(x)&= -\tfrac{1}{2}(\partial_\mu-V_\mu) \xi^\mu(x)-\xi(x)\,,\\
    \delta_{g}\bar{e}(x)&=+ \tfrac{1}{2}(\partial_\mu-V_\mu) \bar{\xi}^\mu(x)-\xi(x)\,,
\end{align}
\end{subequations}
  and
\begin{subequations}\label{eq:boundarygaugetransflv0cld}
    \begin{align}
     \delta_{g}\chi_{\mu}(x)&= \xi_{\mu}(x)-\p_\mu u(x)\,,\\
      \delta_{g}\bar{\chi}_{\mu}(x)&= \bar{\xi}_{\mu}(x)+\p_\mu u(x)\,,\\
      \delta_{g}\chi(x)&=\xi(x)+\tfrac{1}{2}(\p^\mu-V^\mu)\p_\mu u(x)\,.
    \end{align}
\end{subequations}
  Exactly as discussed in Section~\ref{subsec:massless_closed}, it is useful to split the tensor field $e_{\mu\nu}$ into its symmetric and antisymmetric parts, to define the boundary field $l$ according to \eqref{eq:l} and to introduce the field $\phi$ defined in \eqref{eq:physicaldilaton}, which is no longer gauge-invariant. Indeed, by specializing \eqref{eq:gaugetransflv0cld} for $h_{\mu\nu}$, $b_{\mu\nu}$, $l$ and $\phi$, we obtain
  \begin{subequations}
     \begin{align}
        \delta_{g}b_{\mu\nu}(x)&=\partial_{\mu}\lambda_{\nu}(x)-\partial_{\nu}\lambda_{\mu}(x)\,,\\
        \delta_{g}h_{\mu\nu}(x)&=\partial_{\mu}\tau_{\nu}(x)+\partial_{\nu}\tau_{\mu}(x)\,,\\
    \delta_{g}l(y)&=2\tau_z(x)|_{z=0}\,,\\
        \delta_{g}\phi(x)&=\tfrac{1}{2}V_{\mu}\tau^{\mu}(x)\,,
    \end{align}
  \end{subequations}
where we have redefined the gauge parameters \eqref{eq:rotatedgaugeparameter}. 
Under these definitions, the gauge-invariant action \eqref{Stot} becomes
\begin{align}\label{eq:level0closedlineardilaton}
    &S^{\ast}\left(h_{\mu \nu},b_{\mu \nu},\phi;l\right)=\nonumber\\
    &\hspace{0.4cm}=-\frac{\alpha^\prime}{16}\int_{M}d^d x\, e^{-V_\rho x^\rho} \bigg(\tfrac{1}{12}H_{\mu\nu\rho}H^{\mu\nu\rho}\bigg)+\nonumber\\
    &\hspace{0.8cm}-\frac{\alpha^\prime}{16}\int_{M}d^d x\,e^{-V_\rho x^\rho}\bigg(\tfrac{1}{4}\partial_\rho h_{\mu\nu}\partial^\rho h^{\mu\nu}-\tfrac{1}{2}\partial_{\rho}h_{\mu \nu}\partial^\nu h^{\mu\rho}+\tfrac{1}{4}\big(2(\p^\mu-V^\mu)h_{\mu\nu}-\p_\nu \tensor{h}{^{\rho}_{\rho}}\big)\p^\nu \tensor{h}{^{\lambda}_{\lambda}}\nonumber\\
    &\hspace{7.5cm}-2\big((\p^\mu-V^\mu)h_{\mu\nu}-\p_{\nu}\tensor{h}{^{\lambda}_{\lambda}}\big)\p^\nu \phi -4\p_\mu\phi\p^\mu\phi\bigg)+\nonumber\\
    &\hspace{0.8cm}+\frac{\alpha^\prime}{16}\int_{\partial M}d^{d-1} y\,e^{-V_c y^c} \bigg[\tfrac{1}{2}h_{az}\partial^{a}\big(\tensor{h}{^{b}_{b}}-h_{zz}-4\phi\big)-\tfrac{1}{2}h^{ab}\p_a\p_b l-\tfrac{1}{2}\p^a l\p_a(\tensor{h}{^{b}_{b}}-4\phi)\nonumber\\
    &\hspace{10cm}-\tfrac{1}{4}V_z h_{zz}^{2}-\tfrac{1}{4}V_z \p_a l\p^a l\bigg]\,,    
\end{align}
\begin{myblock}
    which is in perfect agreement with the GR result \eqref{eq:CGHSsecond}.
\end{myblock}
For clarity, we have omitted the intricate and rich structures associated with the numerous auxiliary fields, which anyway vanish upon solving their corresponding equations of motion. As a consistency check, we observe that the action above reproduces equation \eqref{eq:level0closedactionfinal2} as we take $V_{\mu}\to 0$. In this case, the effect of working in the linear dilaton background manifests
not only in the exponential prefactor. Indeed, there are additional terms that play a crucial
role in restoring gauge invariance. Interestingly, and in contrast to the free boson case, the
boundary mode $l$ appears not only linearly in the action but also through a quadratic term
proportional to the orthogonal component of the background charge.


\section{Discussion and conclusions}\label{sec:disc}

In this paper we have constructed a gauge-invariant kinetic term for open and closed bosonic SFT for a flat target space with boundary which is given by \eqref{Stot}. Gauge invariance is achieved by promoting the gauge parameter at the boundary to a dynamical boundary string field whose gauge transformation compensates the effect of the boundary. The simplicity of this action and, especially, the simplicity of the way it achieves gauge invariance is quite surprising, especially considering the rather complicated tour we took in order to arrive at it (see \cite{paper1} and appendix \ref{app:equiv}). This can be further substantiated by observing that \eqref{Stot} can be also written in the form \eqref{eq:StotGI} as
\begin{align}
S_{\rm tot}(\Psi,\chi)=\frac12\w{\Big(}(\Psi-Q\chi),(\Theta Q-\delta\Gamma^*) \, (\Psi-Q\chi){\Big)}\,, \label{Stot-new}
\end{align}
where gauge invariance is now manifest. 
This in particular implies that our construction allows for  gauge-invariant  deformations given by a large class of functionals of $\Psi-Q \chi$. However, since the action has to reduce to $\frac12\w(\Psi,Q\Psi)$ when the boundary is sent to infinity and $\Psi$ is taken with sufficient fall-off, these deformations should be necessarily supported at the boundary and have the generic form
\begin{align}
\Delta S=\frac12\w{\Big(}(\Psi-Q\chi),\Delta \, (\Psi-Q\chi){\Big)}\,,\label{eq:DeltaS}
\end{align}
where $\Delta$ is BPZ odd, ghost number 1, localized at the boundary (which means containing $\delta$) and, importantly, without normal derivatives $\p_\bot$ so as not to spoil the well-definiteness of the variational principle ensured by the $c\del_\perp$ in $\Gamma^*$, \eqref{eq:GammaDecomp}. 
These boundary deformations of the action are in one to one correspondence with the freedom in choosing the operator $\gamma^*$ defining the action (see \eqref{eq:gammaDecomp} and \eqref{gammaV}). Indeed, the difference in actions given by two choices $\gamma_{\rm new}^*$ and $\gamma_{\rm old}^*$ of this operator is equal to $\Delta S$ of the form \eqref{eq:DeltaS} with 
$$
\Delta=\delta\,(\gamma_{\rm new}^*-\gamma_{\rm old}^*)=-\delta\,(\gamma_{\rm new}-\gamma_{\rm old})=-\Delta^*\,.
$$
In the explicit examples of this paper we have analyzed the `canonical' case where $\gamma=\gamma^*$ which seems to give a minimal spacetime action, since $\gamma^*$ drops out from \eqref{eq:StotGI} as soon as it is made BPZ-even. It would be interesting to explore this space of boundary deformations and to separate trivial ones (arising  from field redefinitions of the boundary degrees of freedom) from possible physical ones. However doing this at this stage, where we only have control of the free theory, is most probably premature and not very useful as it is to be expected that the consistency of interactions will 
affect this space of deformations in a way which is difficult to envisage at the moment. In other words it is possible that, just like infinitesimal marginal deformations may fail to be exactly marginal, not all choices of $\gamma^*$ will be compatible with adding interactions. This is definitely something to explore in the future.
  
Another point which we would like to clarify in the future is the case of generic curved boundaries. Although in Section \ref{sec:examples} we have only discussed examples of theories on a half-space with a flat boundary, 
it appears
that the action \eqref{Stot} satisfies all consistency criteria regardless of the shape of the boundary. At the same time, in standard gravity with GHY term (which is a subsector of the closed-string theory), it is known that boundary tadpoles emerge when the extrinsic curvature of the boundary $K_{ab}$ fails to satisfy the boundary equation of motion for the metric, namely (see \cite{paper1} for details)
 \begin{align}
 K_{ab}-h_{ab}K\neq0\,.
 \end{align}
Such tadpoles would generally be expected to play a crucial role in restoring gauge-invariance of the action once the interactions are turned on: the free gauge transformation of the kinetic term would be expected to mix with the interacting part of the gauge transformation of the tadpole. One should therefore investigate in what sense \eqref{Stot} is capable of describing free string on a manifold with a generic boundary as it apparently does not need any tadpoles or interactions to restore gauge invariance. For instance, it is possible that upon expanding \eqref{Stot} in component fields for a generic boundary, terms proportional to boundary curvature will appear which will force some combinations of the bulk fields to vanish at the boundary through constraints. Hence, by `dynamically' restricting the bulk field configurations at the boundary in this way, the theory could in principle ensure that a gauge-invariant quadratic action can be written down for a curved boundary even though on general grounds, the presence of a boundary tadpole would be expected (such as in the above-mentioned case of pure gravity). In any case, it is definitely to be expected that in a complete treatment of the case with a generic boundary, interactions should play a fundamental role.


The most pressing question is thus the understanding of string interactions in the presence of a spacetime boundary. In searching for the appropriate construction, it is important to take home the lesson that gauge-invariance should not be taken as the only consistency requirement. In our analysis of the free case, our guiding principles importantly included the requirement of having only first derivatives in the bulk action and no normal derivatives in the boundary action, so as to have a variational principle which gives a well-posed boundary-value problem. 
From this point of view, it will be crucial to understand how this criterion should be adapted when the string coupling constant is turned on, with the consequent emergence of non-local string interactions. 
One possible strategy would be to analyze the problem order-by-order in the non-locality parameter ($\alpha^\prime$) and investigate if one can add suitable boundary terms or possibly make field redefinitions so as to ensure that the variational principle gives a well-posed boundary value problem up to given order in $\alpha^\prime$ (see \cite{Moeller:2002vx,Erbin:2021hkf} for an analysis of the initial value problem in a non-local tachyon EFT).
To this end, one may also find it instructive to revisit the structurally similar case of the non-commutative Chern-Simons theory on a 3-manifold with a boundary, where non-locality arises due to the appearance of the Moyal star product\cite{Seiberg:1999vs,Susskind:2001fb,Pinzul:2001qh}. We are currently working on these and other related  aspects \cite{int} and we will report on our findings in the near future.

\section*{Acknowledgments}
We thank Ted Erler, Martin Schnabl and Jaroslav Scheinpflug for useful discussions. We also thank the participants of the Workshop on String Field Theory and Related Aspects (17-28 March 2025, IGAP, Trieste) for illuminating comments and interactions. The work of CM and AR is partially supported by the MIUR PRIN contract 2020KR4KN2 ``String Theory as a bridge between Gauge Theories and Quantum Gravity'' and by the INFN project ST$\&$FI ``String Theory and Fundamental Interactions''. The work of JV is supported by the ERC Starting Grant 853507.

\appendix


\section{Equivalent forms of the action}\label{app:equiv}

In this appendix we would like to  connect with the approach we followed in the previous paper \cite{paper1} where we used constrained boundary fields having the same ghost number as the bulk field $\Psi$. We will present two actions which we found prior to the discovery of  \eqref{Stot} and which are both closely related to it. We do this partly for keeping track of the route that brought us to \eqref{Stot} and also because these alternative forms may turn useful in the future. 

\subsection{First alternative action}\label{app:equiv-1}
Since in \eqref{Stot} $\chi$ only appears  as $Q\chi$, we can define a new boundary string field $\Sigma$ such that 
\begin{align}
\Sigma=Q\chi\,,\label{Sigma-Q}
\end{align}
so that $\Sigma$ formally gauge-transforms in the same way as $\Psi$ does, namely
\begin{align}
\delta_g\Sigma=Q\Lambda\,.
\end{align}
We can now re-write \eqref{Stot} as
\begin{align}
 S^{(1)}(\Psi, \Sigma)=\frac12 \w(\Psi, (\Theta Q-\delta \Gamma^*)\Psi)+\w(\Psi,\delta\Gamma^* \Sigma)-\frac12\w\left(\Sigma,\delta\Gamma^*\,\Sigma\right)
\end{align}
and we can simply conclude that the gauge-variation is given by
\begin{align}
\delta_g S^{(1)}=-\frac12\w\left(\Lambda,B\,Q\Sigma\right).
\end{align}
This is zero if we assume that $Q\Sigma=0$ which is obviously implied by \eqref{Sigma-Q}.

In the case of the string propagating on the half-space $M=\mathbb{R}_+\times \mathbb{R}^{D-1}$ (see section \ref{sec:examples} for details), the string field $\Sigma$ is understood as a collection of boundary fields 
\begin{align}
\Sigma(x)=\Sigma_1(y)+z\Sigma_2(y)+\frac12 z^2 \Sigma_3(y)+O(z^3)\,.
\end{align}
It is not difficult to realize  that the action only depends on $\Sigma_1$ and $\Sigma_2$, but  still $\Sigma_3$ is needed to realize the constraint $Q\Sigma=0$ which in terms of the boundary fields $\Sigma_i$ reads
 \begin{align}
 \left(\begin{matrix} \tilde Q&-\Omega&-c\\0&\tilde Q&-\Omega\\0&0&\tilde Q    \end{matrix}\right)\left(\begin{matrix}\Sigma_1\\  \Sigma_2\\ \Sigma_{3}\end{matrix}\right) =0\,.\label{Q-constr}
 \end{align}
Explicitly the action in half-space takes the form
 \begin{align}
 S^{(1)}(\Psi;\Sigma_1,\Sigma_2)&=\frac12 \w_M\!\left(\Psi,Q\Psi\right)-\frac12 \w_{\del M}\!\left(\Psi,c\del_z\Psi\right)+\0\\
 &\hspace{1cm}+\frac12\w_{\del M}\left(\Psi,\Omega\Sigma_1+2 c \Sigma_2\right)+\w_{\del M}\left(\Sigma_1,c \Sigma_2\right)\,.
 \end{align}
 The gauge transformations of the $\Sigma_i$'s are simply given 
 \begin{align}
 \delta_g\left(\begin{matrix}\Sigma_1\\ \Sigma_2\\ \Sigma_3     \end{matrix}\right)= \left(\begin{matrix} \tilde Q&-\Omega&-c\\0&\tilde Q&-\Omega\\0&0&\tilde Q    \end{matrix}\right)\left(\begin{matrix}\Lambda\\  \Lambda'\\ \Lambda''\end{matrix}\right) \,,
 \end{align}
 and are obviously compatible with the (gauge-invariant) constraint \eqref{Q-constr} thanks to
 \begin{align}
  \left(\begin{matrix} \tilde Q&-\Omega&-c\\0&\tilde Q&-\Omega\\0&0&\tilde Q    \end{matrix}\right)^2=0\quad \longleftrightarrow\quad Q^2=0\,.
 \end{align}
In principle this action is  more general than \eqref{Stot}, which corresponds to $\Sigma$ being $Q$-exact and not just $Q$-closed. It is not clear to us at the moment whether a cohomological non-trivial $\Sigma$ could play some non-trivial role and in all concrete examples we have analyzed: so far we have always found that \eqref{Stot} is fully sufficient for describing a satisfactory free action for all string modes.

\subsection{Second alternative action}
The last action we would like to mention is infact  the first one which we found and is structurally more similar to the action we used in \cite{paper1}. Again,  we present it  in the half-space $M=\mathbb{R}_+\times \mathbb{R}^{D-1}$,  in the conventions of section \ref{sec:examples}
 \begin{align}
 S^{(2)}(\Psi;\tilde\Sigma_0,\tilde\Sigma_1)&=\frac12 \w_M\!\left(\Psi,Q\Psi\right)-\frac12 \w_{\del M}\!\left(\Psi,c\del_z\Psi\right)+\0\\
 &\hspace{1cm}-\frac12\w_{\del M}\left(\Psi,\Omega\tilde\Sigma_1-2\tilde Q \tilde\Sigma_0\right)+\w_{\del M}\left(\tilde\Sigma_1,\tilde Q \tilde\Sigma_0\right)\,.
 \end{align}
Similarly to the  action described in the previous subsection \ref{app:equiv-1}, the boundary fields $\tilde\Sigma_{1,0}(y)$ are constrained by the relation
 \begin{align}
 \left(\begin{matrix} c&\Omega&-\tilde Q\\0&c&\Omega\\0&0&c    \end{matrix}\right)\left(\begin{matrix}\tilde\Sigma_1\\ \tilde\Sigma_0\\ \tilde\Sigma_{-1}\end{matrix}\right)\equiv \boldsymbol C\boldsymbol \Sigma=0\,, \label{C-constr}
 \end{align}
where $\tilde\Sigma_{-1}$ is a generic boundary field with mass dimension $-1$, which decouples from the action, but which is needed to express the constraint \eqref{C-constr}. Then the action is invariant under the gauge transformation
 \begin{subequations}
 \begin{align}
 \delta_g \Psi&=Q\Lambda\,,\\
 \delta_g \boldsymbol\Sigma&= \boldsymbol C\boldsymbol\Lambda\,,
 \end{align}
 \end{subequations}
after assuming the (gauge-invariant) constraint  \eqref{C-constr} and using the nilpotency relation
 \begin{align}
 \boldsymbol C^2=0\quad \longleftrightarrow \quad Q^2=0\,,
 \end{align}
which is easy to verify.
 
Although it does not look so at first sight, this action is also very closely related to $S_{\rm tot}$ \eqref{Stot} and is in fact equivalent to it. The key for understanding this is to realize that, in this particular case, the constraints \eqref{C-constr} can be explicitly solved by decomposing the $\Sigma_{1,0,-1}$ boundary fields using the ghost zero-modes basic relation
 \begin{align}
 bc+cb=1\,,
 \end{align}
 where $b=b_0$ for the open string and $b=b_0+\bar b_0$ for the closed string. Then, in full generality we can write
 \begin{align}
 \left(\begin{matrix}\tilde\Sigma_1\\ \tilde\Sigma_0\\ \tilde\Sigma_{-1}\end{matrix}\right)= \left(\begin{matrix}\sigma_1+c\, \eta_1\\ \sigma_0+c\, \eta_0\\ \sigma_{-1}+c\, \eta_{-1}\end{matrix}\right),
 \end{align}
 where
 \begin{align}
 b\,\sigma_i=b\,\eta_i=0\,.
 \end{align}
 It is not difficult to see that the constraints \eqref{C-constr} can be uniquely solved by expressing the $\sigma_i$'s in terms of the $\eta_j$'s. One then obtains
 \begin{align}
 \left(\begin{matrix}\tilde\Sigma_1\\ \tilde\Sigma_0\\ \tilde\Sigma_{-1}\end{matrix}\right)= \left(\begin{matrix}-bc \tilde Q \,\eta_{-1}+\Omega\,\eta_0+c\, \eta_1\\ \Omega\,\eta_{-1}+c\, \eta_0\\c\, \eta_{-1}\end{matrix}\right)\,.\label{Sigma-eta}
 \end{align}
The action $S^{(2)}$ can thus be expressed as
\begin{align}
S^{(2)}(\Psi,\tilde\Sigma_i)\longrightarrow S^{(2)}(\Psi,\tilde\Sigma_i(\eta_j))\equiv \tilde S^{(2)}(\Psi;\eta_{-1},\eta_0,\eta_1)\,.
\end{align}
We can find the gauge transformations of the $\eta$ fields by realizing from \eqref{Sigma-eta} that we can write  
\begin{align}
\eta_i=b\,\tilde\Sigma_i\,,
\end{align}
which implies
\begin{subequations}
\begin{align}
\delta_g \eta_{-1}&=b(c\Lambda)\,,\\
\delta_g \eta_{0}&=b(c\Lambda'+\Omega \Lambda)\,,\label{gauge-eta}\\
\delta_g \eta_{1}&=b(c\Lambda''+\Omega \Lambda'-\tilde Q\Lambda)\,.
\end{align}
\end{subequations}
Notice that the $\eta$ fields have degree $(-1)$ and opposite grassmannality wrt the bulk string field $\Psi$. This already suggests of a possible relation between the three $\eta$ fields in $\tilde S^{(2)}$ and the three $\chi$ fields of $S_{\rm tot}$ \eqref{Stot}, whose $\Lambda$-gauge transformations \eqref{chi-gauge} are however rather different from \eqref{gauge-eta}. Moreover, the $\eta$ fields are annihilated by $b$ while this is not true for the $\chi$ fields. The key point here is that $\chi$ has an extra  gauge symmetry given by $Q\Upsilon$  \eqref{gauge2} and this can be used to put $\chi$ in Siegel gauge without touching the $\Lambda$ gauge-symmetry. To do this we use the $b$-$c$ decomposition 
 \begin{align}
 \chi_i&=\beta_i+c\gamma_i
 \,,
 \end{align}
 (where $b\beta_i=b\gamma_i=0$) and search for a gauge parameter  \begin{align}
 \Upsilon(z)=\frac 12 z^2 \Upsilon_{-1}+\frac1{3!} z^3\Upsilon_0+\frac1{4!} z^4 \Upsilon_1\,,
 \end{align}
 to put $\chi_{{-1,0,1}}$ in Siegel gauge.\footnote{Other $z$ powers in $\Upsilon$ are not needed for bringing $\chi_{-1,0,1}$ in Siegel gauge.}

By explicitly computing $Q\Upsilon$ we get
\begin{subequations}
\begin{align}
\delta_{_\Upsilon} \chi_{-1}&=-c\,\Upsilon_{-1}\,,\\
\delta_{_\Upsilon}\chi_{0}&=-c\,\Upsilon_{0}-\Omega\,\Upsilon_{-1}\,,\\
\delta_{_\Upsilon} \chi_{1}&=-c\,\Upsilon_{1}-\Omega\,\Upsilon_{0}+\tilde Q\, \Upsilon_{-1}\,.
\end{align}
\end{subequations}
 We can then easily determine $\Upsilon_{-1,0,1}$ so that 
 \begin{align}
\chi\longrightarrow\chi+Q\Upsilon=\hat\beta+c\hat\gamma,\quad \textrm{with } \,\hat\gamma=0\,.
 \end{align}
  By simple matching, we end up with
 \begin{subequations}
 \begin{align}
\hat\beta_{-1}&=\beta_{-1}\,,\\
\hat\beta_0&=\beta_{0}-\Omega\,\gamma_{-1}\,,\\
\hat\beta_1&=\beta_{1}-\Omega\,\gamma_{0}+bc\tilde Q\,\gamma_{-1}\,.
 \end{align}
 \end{subequations}
 These explicit relations allow  to determine how the $\Lambda$-gauge transformation acts on the new variables $\hat\beta_i$ ,
  by tracing back its action on the original $\beta=bc\,\chi$ and $\gamma=b\chi$ variables. This gives
 \begin{subequations}
     \label{gauge-beta}
 \begin{align}
 \delta_\Lambda \hat\beta_{-1}&=bc\,\Lambda\,,\\
  \delta_\Lambda \hat\beta_{0}&=b(c\,\Lambda'+\Omega\,\Lambda)\,,\\
  \delta_\Lambda \hat\beta_{1}&=b(c\,\Lambda''+\Omega\,\Lambda'-\tilde Qcb\,\Lambda)\,.
 \end{align}
 \end{subequations}
 These gauge transformations are now rather similar to the  gauge transformations of the $\eta$ fields \eqref{gauge-eta} and become identical after the field redefinition
 \begin{subequations}
 \begin{align}
 \hat\beta_{-1}&=\eta_{-1}\,,\\
\hat\beta_0&=\eta_{0}\,,\\
\hat\beta_1&=\eta_{1}-b\tilde Q\,\eta_{-1}\,.
\end{align}
\end{subequations}
This gives the final relation between the $\eta$ fields and the $\chi$ fields and establishes that
\begin{align}
\tilde S^{(2)}(\Psi, \eta)=S_{\rm tot}(\Psi, \hat\beta(\eta))\,.
\end{align}


\endgroup

\end{document}